\documentclass[aps,prc,twocolumn,groupaddress,showpacs,10pt,dvipsnames]{revtex4-1}

\usepackage[utf8]{inputenc}
\usepackage{amsfonts}
\usepackage{amsmath}
\usepackage{amssymb}
\usepackage{bm}
\usepackage[pdftex]{graphicx}
\usepackage[dvipsnames]{xcolor}
\usepackage[normalem]{ulem}
\usepackage{comment}
\usepackage{multirow}

\newcommand{\tj}[6]{ \begin{pmatrix}
  #1 & #2 & #3 \\
  #4 & #5 & #6
 \end{pmatrix}}
 
 \newcommand{\sj}[6]{ \begin{Bmatrix}
  #1 & #2 & #3 \\
  #4 & #5 & #6
 \end{Bmatrix}}

\begin{document}

% Use the \preprint command to place your local institutional report
% number in the upper righthand corner of the title page in preprint mode.
% Multiple \preprint commands are allowed.
% Use the 'preprintnumbers' class option to override journal defaults
% to display numbers if necessary
%\preprint{}

%Title of paper
\title{Opening angle and dineutron correlations in knockout reactions with Borromean two-neutron halo nuclei} 

% repeat the \author .. \affiliation  etc. as needed
% \email, \thanks, \homepage, \altaffiliation all apply to the current
% author. Explanatory text should go in the []'s, actual e-mail
% address or url should go in the {}'s for \email and \homepage.
% Please use the appropriate macro foreach each type of information

% \affiliation command applies to all authors since the last
% \affiliation command. The \affiliation command should follow the
% other information
% \affiliation can be followed by \email, \homepage, \thanks as well.

%\email[]{Your e-mail address}
%\homepage[]{Your web page}
%\thanks{}
%\altaffiliation{}

\author{J. Casal}
\email{jcasal@us.es}
\affiliation{Departamento de F\'{\i}sica At\'omica, Molecular y Nuclear, Facultad de F\'{\i}sica, Universidad de Sevilla, Apartado 1065, E-41080 Sevilla, Spain} 
\author{M. Gómez-Ramos}
\email{mgomez40@us.es}
\affiliation{Departamento de F\'{\i}sica At\'omica, Molecular y Nuclear, Facultad de F\'{\i}sica, Universidad de Sevilla, Apartado 1065, E-41080 Sevilla, Spain}

%Collaboration name if desired (requires use of superscriptaddress
%option in \documentclass). \noaffiliation is required (may also be
%used with the \author command).
%\collaboration can be followed by \email, \homepage, \thanks as well.
%\collaboration{}
%\noaffiliation

\date{\today}

\begin{abstract}
 \begin{description}
  \item[Background] Knockout reactions with proton targets provide an invaluable tool to access the properties of two-neutron halo nuclei. Recently, experimental results for the average opening angle as a function of the intrinsic neutron momentum in $^{11}$Li have shown a localization of dineutron correlations on the nucleus surface.
  \item[Purpose] Study the model dependence and the effect of distortion and absorption on the opening angle distributions to assess the reliability of this observable to extract properties of Borromean two-neutron halo nuclei.
  \item[Method] A quasifree sudden model is used to describe the knockout process, where absorption effects are modeled by the eikonal $S$-matrix between the proton target and the core of the Borromean nucleus. Final states in momentum space are built within a three-body model for the projectile, which enables the description of momenta and opening angle distributions. 
  \item[Results] A strong dependence on absorption effects is found for the opening angle at large intrinsic momenta, while the region of lower momenta is mostly insensitive to them. Reasonable agreement with the available data is obtained for $^{11}$Li at low momenta with weights for $s$ and $p$ waves different from those previously reported, showing a model dependence in their extraction. For $^{19}$B, test calculations show marked sensitivity to small $p$-wave components.
  \item[Conclusions] The opening angle for $(p,pn)$ knockout reactions on Borromean nuclei at small intrinsic momenta is a reliable observable mostly sensitive to the structure of the Borromean nucleus. For larger momenta, the reaction mechanism leads to a larger distortion of the distribution. In the case of nuclei with small components of opposite parity to the dominant ones, this observable can be used to explore them. The relation between dineutron in coordinate space and opening angle in momentum space is found to be model-dependent.
 \end{description} 
\end{abstract}

% insert suggested PACS numbers in braces on next line
%\pacs{ }%21.45.-v, 26.20.-f, 26.30.-k,27.20.+n}
% insert suggested keywords - APS authors don't need to do this
%\keywords{}

%\maketitle must follow title, authors, abstract, \pacs, and \keywords
\maketitle

% body of paper here - Use proper section commands
% References should be done using the \cite, \ref, and \label commands
  %\section{}
% Put \label in argument of \section for cross-referencing
%\section{\label{}}
  %\subsection{}
  %\subsubsection{}

\section{Introduction}
\label{sec:intro}
Two-neutron halo nuclei along the neutron dripline are at the focus of our efforts to understand the limits of nuclear stability. Since the first observation of an abnormally large interaction cross section for $^{11}$Li~\cite{tanihata85}, the topic has driven enormous experimental and theoretical endeavours~\cite{tanihata2013}. The term halo refers to a diffuse matter distribution corresponding to the valence neutrons, which are loosely bound and explore distances far from the more compact core~\cite{hansen87,riisager94}. The structure of $\text{core}+n+n$ two-neutron halo nuclei is usually called Borromean~\cite{zhukov93,nielsen01}, where the binary subsystems $\text{core}+n$ and $n+n$ are unbound. It is then clear that the correlations between the valence neutrons are essential in binding the system~\cite{zhukov93,esbensen97,hagino05}. These correlations favor a strongly localized two-neutron structure, also referred to as dineutron configurations, which is enhanced by a large mixing between different-parity orbitals~\cite{catara84}. In $^{11}$Li, for instance, three-body calculations show that the halo wavefunction is very much determined by the mixing between $s$- and $p$-wave states in the low-lying spectrum of the unbound $^{10}$Li system~\cite{hagino05,kikuchi2013,GomezRamos2017plb}, and a dominant dineutron peak is obtained in the corresponding two-neutron density. A similar situation has been recently explored for the heavier two-neutron halo $^{29}$F, where the mixing between intruder $p_{3/2}$ components with standard-order $d_{3/2}$ enhances the dineutron configuration and the size of the halo~\cite{bagchi2020,singh2020}. In the case of two-neutron halo nuclei without a strong mixing between different-parity components, such as $^{6}$He or $^{19}$B, the dineutron is less pronounced~\cite{zhukov93,cook2020,casal2020b19}. 

Different techniques have been employed to investigate the correlations between the halo neutrons experimentally. The angle between the two valence neutrons in $^{11}$Li was estimated from a Coulomb breakup measurement, using the link between the extracted $E1$ strength into the continuum and the so-called cluster sum rule in a $\text{core}+n+n$ model, and assuming an inert core~\cite{Nakamura06}. The average angle obtained was $\langle \theta_{nn}\rangle=48^{+14}_{-18}$ degrees, well below the value of 90 degrees expected for a no-correlation scenario~\cite{Bertsch91,Esbensen92}. This estimation was refined in a subsequent theoretical work, pointing towards a larger $\langle \theta_{nn} \rangle$ value, but always compatible with a correlated pair in coordinate space~\cite{Bertulani07}. It is worth noting, however, that the simple relation between the cluster sum rule for dipole transitions and the geometrical configuration of the two-neutron halo is model dependent, and effects such as core excitations can modify the results~\cite{kikuchi2013}. 
An alternative way to access these correlations is provided by knockout reactions~\cite{Sim99}. In Ref.~\cite{Sim07}, the neutron-knockout from $^{11}$Li and $^{14}$Be on a carbon target was reported and analyzed in terms of the mixing between different components in the projectile wavefunction. For $^{11}$Li, an asymmetric angular distribution was obtained, favoring a large opening angle ($> 90$ deg.) between the neutrons in momentum space, which translates to a small angle in coordinate space, and therefore small distance between neutrons: a dineutron structure.

Recently, the dineutron correlation in $^{11}$Li has been further studied with the measurement of the neutron-knockout on a proton target at intermediate energies~\cite{Kubota2020}, and analyzed in terms of the quasi-free eikonal sudden model from Ref.~\cite{Kikuchi2016}. The missing momentum (i.e., the intrinsic momentum of the knocked-out neutron) distribution was described in terms of $s$-, $p$- and $d$-waves in the $^{11}$Li ground state, and their relative weights were estimated from the fitting of the data. From the asymmetry in the opening angle and its dependence on the missing momentum, explored for the first time, to our knowledge, their conclusions indicate that the dineutron correlation is localized on the surface of the $^{11}$Li halo. The sensitivity to the reaction mechanism and to the degree of mixing between different-parity states, however, was not discussed in detail and may play a role in the extracted conclusions.

In this work we explore this new observable, the opening angle as a function of the missing momentum, and analyze its sensitivity to the reaction mechanism and to the structure of the projectile, described through a three-body model, in order to determine which properties of the Borromean nucleus can be extracted from its analysis. In Section \ref{sec:theory} we rederive the expressions presented in \cite{Kikuchi2016}, modified to suit a specific three-body framework used to describe Borromean nuclei. In Section~\ref{sec:results}, results for the nucleon removal $(p,pn)$ reaction on $^{11}$Li and $^{19}$B are presented, focusing on the effect of the reaction mechanism and their nuclear structure. Finally, in Section~\ref{sec:conclusions}, the conclusions and outlook of this work are summarized.

\section{Theoretical framework}
\label{sec:theory}
We will describe the momentum and angular distributions in the knockout of a neutron from two-neutron halo nuclei by following the theoretical model in Ref.~\cite{Kikuchi2016}.%, which has been recently employed to analyze the angular correlations leading to the surface localization of the dineutron in $^{11}$Li~\cite{Kubota2020}. 

\begin{figure}[t]
 \centering
 \includegraphics[width=0.5\linewidth]{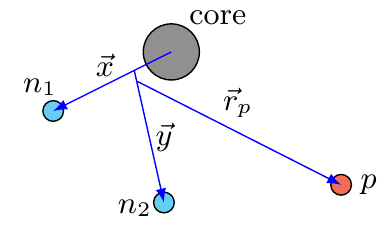}
 \caption{Schematic representation of the neutron-knockout process from the halo of a $\text{core}+n+n$ system.}
 \label{fig:knockout}
\end{figure}

Assuming a quasi-free regime, the transition potential for the $(p,pn)$ process involves the interaction between the target proton and the knocked-out neutron. In a zero-range approximation, we can write the T-matrix as
\begin{equation}
        T= \langle\Psi^{(-)}|V_{np}\delta(\boldsymbol{r}_p-\boldsymbol{r}_2)|\chi_p \Phi_{gs}\rangle
\end{equation}
where $\Psi^{(-)}$ is the final state wavefunction, $\chi_p$ is the distorted wave for the proton, and $\boldsymbol{r}_p$ and $\boldsymbol{r}_2$ denote the position of the proton and removed neutron with respect to the center of mass of the projectile. Here, $\Phi_{gs}$ corresponds to the ground-state wavefunction of the bound Borromean ($\text{core}+n+n$) nucleus. Using an eikonal sudden approach, the proton distorted wave will transfer a momentum $\boldsymbol{q}$ to the removed nucleon and can be written as
\begin{equation}
    \label{eq:protonDW}
    \chi_p=S(\boldsymbol{b}_p)e^{i\boldsymbol{q}\cdot \boldsymbol{r}_p},
\end{equation}
with $\boldsymbol{b}_p$ being the impact parameter of the proton and $S(\boldsymbol{b}_p)$ the usual eikonal $S$-matrix,
\begin{equation}
    S(\boldsymbol{b}_p)=\exp\left[{-\frac{i}{\hbar v}\int_{-\infty}^\infty V_{pC}(\boldsymbol{b}_p,z_p) \mathrm{d}z_p}\right].
\end{equation}
Thus we get
\begin{equation}
        T= V_{np} \langle \Psi^{3.b.(-)}|\delta(\boldsymbol{r}_p-\boldsymbol{r}_2)|S(\boldsymbol{b}_p) e^{i\boldsymbol{q}\cdot \boldsymbol{r}_p} \Phi_{gs}\rangle,
\end{equation}
where we will assume that the final state of the two neutrons and the residual core $\Psi^{3.b.(-)}$ is not distorted by the proton after the collision, except for the transferred momentum $\boldsymbol{q}$. The knocked-out neutron will have a momentum $\boldsymbol{k}^f_y=\boldsymbol{k}_y+\boldsymbol{q}$ in the final state, where $\boldsymbol{k}_y$ is the ``original'' momentum the neutron had ``inside'' the Borromean projectile. Since the mass for the core is much larger than that of the valence particles, we can approximate $\boldsymbol{r}_2$ by $\boldsymbol{y}$, the distance between the removed neutron and the center of mass of the remaining $\text{core}$-$n$ subsystem. Moreover the transferred momentum at high energies in quasi-free kinematics will be large, so the wavefunction of the removed nucleon will be similar to a free plane wave. Therefore (ignoring spins) the final state of the two neutrons and the core can be expressed as
\begin{equation}
    \Psi^{3.b.(-)}_{\text{no spins}}=\phi_{\rm c\text{-}n}(\boldsymbol{k}_x,\boldsymbol{x}) e^{i(\boldsymbol{k}_y+\boldsymbol{q}) \cdot \boldsymbol{y}},
\end{equation}
where $\boldsymbol{x}$ is the coordinate between the core and the neutron that is not removed in the $(p,pn)$ reaction, so that $\phi_{\rm c\text{-}n}(\boldsymbol{k}_x,\boldsymbol{x})$ is the continuum wavefunction for the core-neutron system corresponding to an asymptotic momentum $\boldsymbol{k}_x$. The distances $\boldsymbol{x}$ and $\boldsymbol{y}$ can be easily related to the Jacobi coordinates~\cite{zhukov93}, see Fig.~\ref{fig:knockout} . Now the application of the zero-range approximation for the $V_{pn}$ interaction, and the inclusion of the coupling of all relevant spins in the final state, yields for the T-matrix
\begin{equation}
        T\propto \langle \phi_{\rm c\text{-}n}(\boldsymbol{k}_x,\boldsymbol{x}) \otimes e^{i\boldsymbol{k}_y \cdot \boldsymbol{y}}|S(\boldsymbol{b}_y) \Phi_{gs}(\boldsymbol{x},\boldsymbol{y})\rangle.
        \label{eq:fou1}
\end{equation}
As a final simplifying approximation, we will consider that the $S$-matrix dependence on $\boldsymbol{b}_y$ can be approximated by the dependence on the modulus of $y$, as in Ref.~\cite{Kikuchi2016}, 
\begin{equation}
        T\propto \langle \phi_{\rm c\text{-}n}(\boldsymbol{k}_x,\boldsymbol{x}) \otimes e^{i\boldsymbol{k}_y \cdot \boldsymbol{y}}|S(y) \Phi_{gs}(\boldsymbol{x},\boldsymbol{y})\rangle.
        \label{eq:fou2}
\end{equation}
In the previous expression, $S(y)$ introduces distortion and absorption in the $(p,pn)$ quasi-free process, so the interior of the nucleus will have a diminished effect in the final cross section. It is clear that, if no absorption is included ($S=1$) and the $\text{core}+n$ continuum state is replaced by a plane wave in the $\boldsymbol{x}$ coordinate, we recover for $T$ a simple Fourier transform for the ground-state wavefunction $\Phi_{gs}$ of the Borromean projectile. Therefore, in Eq.~(\ref{eq:fou2}) we recognize a distorted Fourier transform. The corresponding cross section, as a function of $\boldsymbol{k}_x,\boldsymbol{k}_y$, matches the result derived in Ref.~\cite{Kikuchi2016}.

\subsection{Three-body wavefunctions}
\label{sec:3b}

The modified Fourier transform introduced in Eq.~(\ref{eq:fou2}) can be written as
\begin{equation}
 \begin{split}
    \Psi^{j\mu}(\boldsymbol{k}_x,\boldsymbol{k}_y) & = \left\langle \phi_{\rm c\text{-}n}(\boldsymbol{k}_x,\boldsymbol{x}) \otimes e^{i\boldsymbol{k}_y \boldsymbol{y}}\big|S(y)\Phi_{\rm g.s.}^{j\mu}(\boldsymbol{x},\boldsymbol{y})\right\rangle \\
    & = \sum_{\eta}w_{\eta}(k_x,k_y) \\
    & \times \left\{\left[\mathcal{Y}_{l_xs}^{j_x}(\Omega_x)\otimes \chi_{I}\right]_{j_1} \otimes \left[Y_{l_y}(\Omega_y)\otimes \kappa_{s}\right]_{j_2}\right\}_{j\mu},
   \end{split}
    \label{eq:fourier1}
\end{equation}
where the label $\eta\equiv\{l_x,j_x,I,j_1,l_y,j_2\}_j$ follows the coupling scheme in Refs.~\cite{Casal2017plb,GomezRamos2017plb}, i.e., $\boldsymbol{j}_x=\boldsymbol{l}_x+\boldsymbol{s}$ is the single-particle angular momentum of a neutron with respect to the core, so that
\begin{equation}
   \mathcal{Y}_{l_xs}^{j_xm_x}(\Omega_x) = [Y_{l_x}(\Omega_x)\otimes\kappa_s]_{j_xm_x},
   \label{eq:2bcoup}
\end{equation}
$j_1$ is the total angular momentum of the core-$n$ system after coupling with the spin $I$ of the core, $j_2$ is the single-particle angular momentum of the remaining neutron, and $\boldsymbol{j}=\boldsymbol{j}_1+\boldsymbol{j}_2$. Note that this wavefunction is expressed in the so-called Jacobi-$Y$ representation, where the conjugated $\boldsymbol{x}$ coordinate connects the core and one valence neutron. 

The momentum functions $w_{\eta}(k_x,k_y)$ in Eq.~(\ref{eq:fourier1}) can be obtained as
\begin{equation}
  \begin{split}
    w_\eta(k_x,k_y)& =(4\pi)^2\sum_{c'}\dfrac{i^{-l_x-l_y}}{k_x}\int dx dy f^{(j_1)}_{c,c'} (k_x,x) \\
    & \times \omega_{c',j_1,l_y,j_2}(x,y) j_{l_y}(k_y y)S(y)y,
  \end{split}
  \label{eq:wfun}
\end{equation}
where $c=\{l_x,j_x,I\}$, so that $\eta=\{c,j_1,l_y,j_2\}$. Here, $f^{(j_1)}_{c,c'} (k_x,x)$ is the radial part of the $\text{core}+n$ continuum state $\phi_\text{c-n}(\boldsymbol{k}_x,\boldsymbol{x})$ for specific entrance and exit channels, $j_{l_y}(k_yy)$ are spherical Bessel functions from the partial-wave expansion of the plane wave, and $\omega_{\eta}(x,y)$ are the radial wavefunctions of $\Phi_{g.s.}(\boldsymbol{x},\boldsymbol{y})$ in Jacobi coordinates. This wavefunction follows the same form of Eq.~(\ref{eq:fourier1}),
\begin{equation}
 \begin{split}
    \Phi_{\rm g.s.}^{j\mu}(\boldsymbol{x},\boldsymbol{y})& =\frac{1}{xy}\sum_{\eta}\omega_{\eta}(x,y)\\
    & \times\left\{\left[\mathcal{Y}_{l_xs}^{j_x}(\widehat{x})\otimes \chi_{I}\right]_{j_1} \otimes \left[Y_{l_y}(\widehat{y})\otimes \kappa_{s}\right]_{j_2}\right\}_{j\mu},
 \end{split}
    \label{eq:wfcoord}
\end{equation}
where the angles now refer to the spatial coordinates. Details on how to construct this coordinate-space wavefunction for $\text{core}+n+n$ nuclei are given, for instance, in Ref.~\cite{casal2020f29} and are summarized in the Appendix.

By introducing the radial overlaps between the three-body ground state and the two-body continuum wavefunctions used in Refs.~\cite{Casal2017plb,GomezRamos2017plb}, 
\begin{equation}
     \xi_{c'}^{\eta}(k_x,y)=\int dx  f^{(j_1)}_{c,c'} (k_x,x) \omega_{c',j_1,l_y,j_2}(x,y),
    \label{eq:overppn}
\end{equation}
the previous expression can be rewritten,
\begin{equation}
   \begin{split}
    & w_\eta(k_x,k_y)  \\ & = (4\pi)^2\sum_{c'}\frac{i^{-l_x-l_y}}{k_x}\int dy  \xi_{c'}^{\eta}(k_x,y) j_{l_y}(k_y y)S(y)y.
    \end{split}
    \label{eq:wfunover}
\end{equation}
Finally, the overlaps for different channels $\{l_x',j_x',I'\}$ corresponding to the same configuration of the knocked-out neutron $\{ly,j_2\}$ can be summed, leading to
\begin{equation}
    \tilde{\xi}_{\eta}(k_x,y)=\sum_{c'} \xi_{c'}^{\eta}(k_x,y),
    \label{eq:mergover}
\end{equation}
so we get
\begin{equation}
   \begin{split}
    & w_\eta(k_x,k_y) \\
    &= (4\pi)^2\frac{i^{-l_x-l_y}}{k_x}\int dy \tilde{\xi}_{\eta}(k_x,y) j_{l_y}(k_y y)S(y)y.
    \end{split}
    \label{eq:wfunmerg}
\end{equation}

\subsection{Momentum and angular distributions}
\label{sec:algebra}
\begin{figure}[t]
 \centering
 \includegraphics[width=0.4\linewidth]{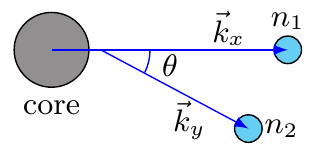}
 \caption{Relative momenta $k_x$ and $k_y$, and opening angle $\theta$ used in the present work to characterize the knockout process.}
 \label{fig:opangle}
\end{figure}

From Eqs.~(\ref{eq:fou2}) and~(\ref{eq:fourier1}), the cross section for the $(p,pn)$ process is obtained as
\begin{equation}
    \sigma\propto \frac{1}{2j+1}\sum_\mu \int d\boldsymbol{k}_xd\boldsymbol{k}_y\left|\Psi\right|^2,
    \label{eq:xsec0}
\end{equation}
where the dependence on $k_x,k_y$ and the angles $\Omega_x,\Omega_y$ is implicit in the wavefunction density. To obtain momentum distributions and the relative angle between the two Jacobi vectors, we can define the $z$ axis in the direction of $\boldsymbol{k}_x$. In that case, $\theta_y=\theta$ and $k_y$ represent, respectively, the opening angle and the missing momentum considered in Ref.~\cite{Kubota2020} and shown in Fig.~\ref{fig:opangle}. Using this condition for the construction of the density, integrating all variables except $k_x,k_y$ and $\theta$ and working out the algebra, we get
\begin{equation}
\begin{split}
    \sigma & \propto  \int dk_xdk_y d\left(\cos{\theta}\right)\sum_{\eta\eta'} w_\eta(k_x,k_y)w^*_{\eta'}(k_x,k_y) C_{\eta\eta'}\\&\times\sum_L D_{\eta\eta'}^{(L)} \tj{l_y}{l_y'}{L}{0}{0}{0}\tj{l_x}{l_x'}{L}{0}{0}{0} P_{L}(\cos\theta),
\end{split}
    \label{eq:49simp}
\end{equation}
where the $C_{\eta\eta'}$ and $D_{\eta\eta'}^{(L)}$ constants are given by products of angular-momentum factors, 6$j$ symbols and real phases. More details are provided in the Appendix. From this expression, the extraction of the opening angle and the momentum distribution is trivial. 
The angular dependence in Eq.~(\ref{eq:49simp}) is given by Legendre polynomials $P_L$ of order $L$ in $\cos\theta$, and we see that $L$ comes from coupling different orbital angular momenta of the three-body wavefunction. Since the Legendre polynomials are symmetric funtions for even $L$ values, it is then clear that asymmetric angular distributions (associated to $nn$ correlations) can only be obtained if different-parity states in the $\text{core}$-$n$ subsystem are combined, which is consistent with traditional literature~\cite{catara84} and with the results discussed for $^{11}$Li in Ref.~\cite{Kubota2020}. This mixing will be enhanced if the different-parity components show a large degree of overlap in $k_y$ and $k_x$.

\section{Results}
\label{sec:results}
\subsection{Application to $^{11}$Li}
\label{sec:application}
We apply the present formalism to the $^{11}\text{Li}(p,pn)^{10}\text{Li}$ reaction in inverse kinematics, for which recent experimental data on the missing momentum and opening angle are available~\cite{Kubota2020}. To compute the relevant distributions, we use as a reference the three-body ($^{9}\text{Li}+n+n$) ground-state wavefunction for $^{11}$Li used in  Refs.~\cite{Casal2017plb,GomezRamos2017plb}. The model includes the spin of $^{9}$Li, giving rise to a splitting of the $2s_{1/2}$ and $1p_{1/2}$ single-particle levels into $1^-,2^-$ and $1^+,2^+$ doublets, whose positions are adjusted with an effective neutron-core potential assuming an inert core. To that aim we employ the potential labeled as P1I in Refs.~\cite{Casal2017plb,GomezRamos2017plb}, which follows the spin-dependent parametrization of Ref.~\cite{Gar03}. For the $nn$ interaction, we adopt the Gogny-Pirres-Tourreil tensor potential~\cite{GPT}.

The $^{11}$Li ground state so obtained is characterized by a dominance of $s$-wave components, and it provides a good description of both low-energy $(p,d)$ angular distributions and quasifree $(p,pn)$ relative-energy spectra. In this work, besides the $s$ and $p$ states considered in Refs.~\cite{Casal2017plb,GomezRamos2017plb}, we include also $1d_{5/2}$ resonances ($4^-,3^-,2^-,1^-$) starting at $E_{\text{core-}n}=4.5$ MeV, which was suggested in Ref.~\cite{Moro2019} to better describe $^{9}\text{Li}(d,p)^{10}$Li spectra. This results in a small $d$-wave contribution to the ground-state wavefunction. A summary of the structure properties of the $^{11}$Li ground state, including the corresponding $^{10}$Li states considered to fit the $\text{core}+n$ potential, are given in Table~\ref{tab:prop}. Details of the three-body calculations, which are based on the hyperspherical-harmonics formalism~\cite{zhukov93}, are given in the Appendix.

\begin{table}[t]
    \centering
    \begin{tabular}{cccccccccc}
      \toprule
        \multicolumn{2}{c}{$a$ (fm)} & & \multicolumn{3}{c}{$E_r$ (MeV)} & & \%$s_{1/2}$ & \%$p_{1/2}$ & \%$d_{5/2}$ \\
        2$^-$ & 1$^-$                & & 1$^+$  & 2$^+$  & 4$^-$         & &              &              &              \\
      \colrule
        -38   &   -                  & & 0.37   & 0.61   & 4.5           & & 63           & 32           & 2            \\
      \botrule
    \end{tabular}
    \caption{Properties of $^{10}$Li (scattering length $a$ or resonance energy $E_r$ for $s$- and $p,d$-wave states, respectively) and $^{11}$Li (partial-wave content in the ground state) with the adopted model.}
    \label{tab:prop}
\end{table}

\begin{figure}[t]
 \centering
 \includegraphics[width=0.9\linewidth]{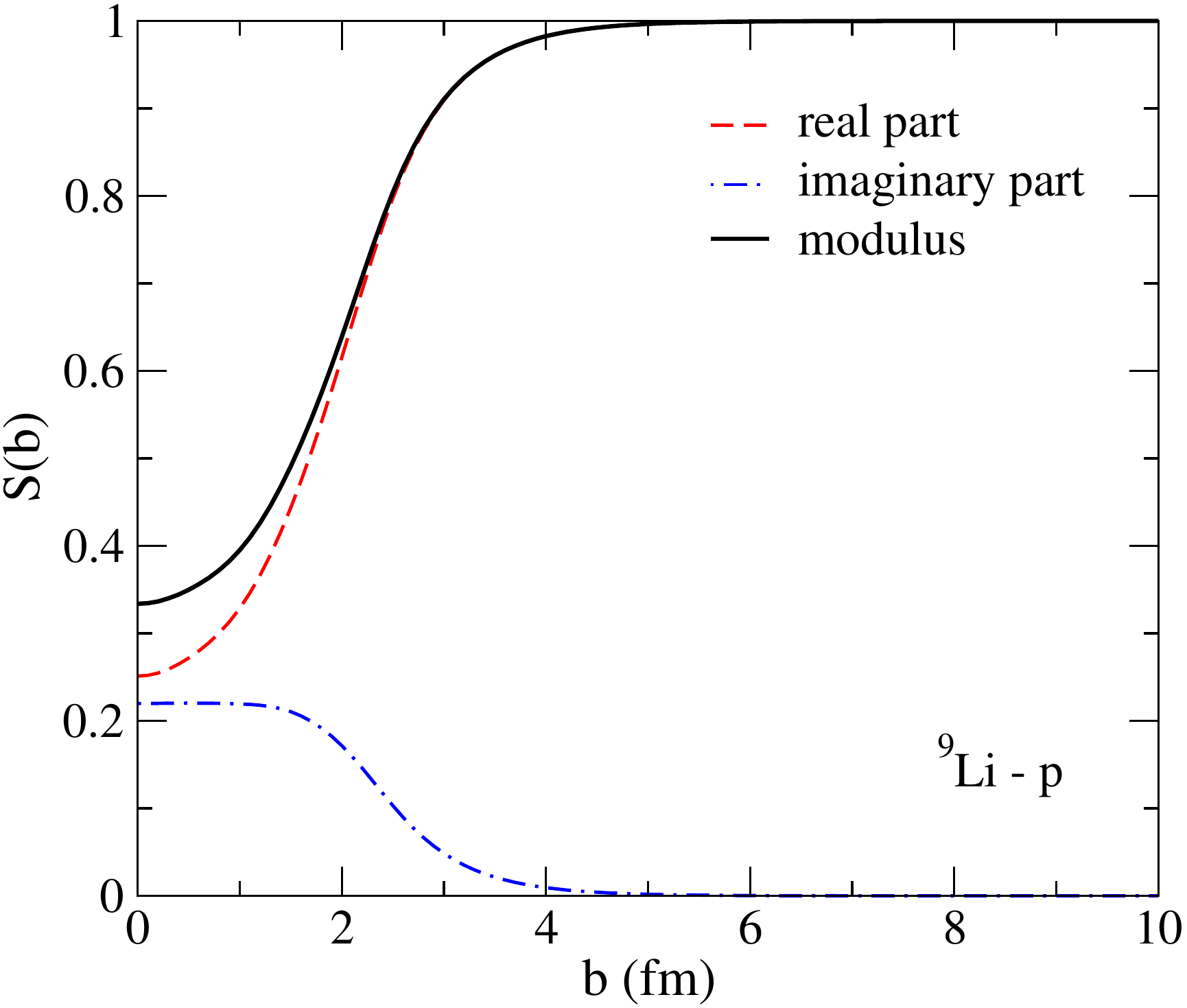}
 \caption{$S$-matrix for $^9$Li-$p$ used in the calculation. It has been computed through the $t\rho$ prescription using for $^9$Li a Hartree-Fock density with the SkX interaction. }
 \label{fig:smat}
\end{figure}

Following Ref.~\cite{Kikuchi2016}, we model the absorption by the proton target through the modulus of an eikonal $S$-matrix between proton and $^9$Li computed using the t$\rho$ prescription, as in many previous nucleon knockout analyses \cite{Gad08,Tos14}, using a Hartree-Fock density with the SkX interaction \cite{Bro98} for $^9$Li. The $S$-matrix is presented in Fig.~\ref{fig:smat}. As can be seen in the figure, it introduces a mild absorptive effect for small distances between proton and core, which also correspond to small distances between neutron and core, due to the zero-range approximation for the $V_{pn}$ interaction.

\begin{figure}[t]
 \centering
 \includegraphics[width=0.95\linewidth]{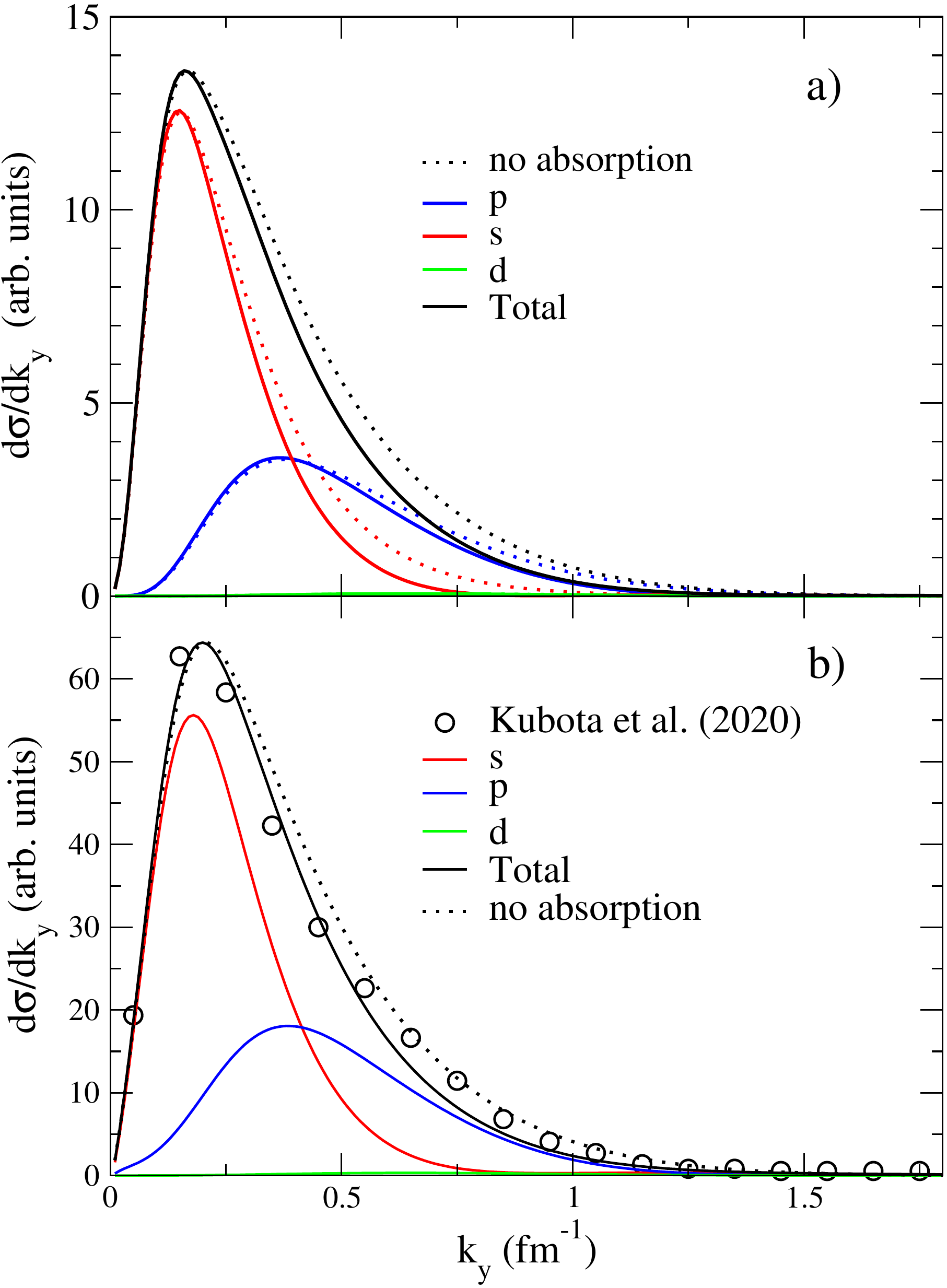}
 \caption{a) Missing momentum distribution for the $^{11}$Li$(p,pn)$ reaction. The different components are presented in solid lines, as well as the contributions setting $S(y)=1$, in dotted lines, rescaled to reproduce the maximum of the distributions with absorption. b) The missing momentum distribution is convoluted with the experimental resolution and compared to the experimental data of \cite{Kubota2020}.}
 \label{fig:kmiss}
\end{figure}

Firstly we present in Fig.~\ref{fig:kmiss} the missing momentum ($k_y$) distribution for the $^{11}$Li$(p,pn)$ reaction. In the top panel we present the momentum distribution and its decomposition in $s$-, $p$- and $d$-wave components, in solid lines. As can be seen in the figure, the $d$-wave has a negligible effect in this observable, due to its small contribution to the ground state of $^{11}$Li. %It should be noted that, for this observable, each component contributes incoherently to the others, so that it is only sensitive to the modulus of $S(y)$, which is related to the absorption of the proton by the core. 
In order to explore the effect of the absorption of the proton, we present as well results nullifying absorption by setting $S(y)=1$, in dotted lines. The effect is moderate at best, narrowing the distributions somehow, and having almost no effect at small $k_y$. This is expected, since the removed neutron is a halo neutron, so its collision with the proton takes place at a large distance from the $^9$Li core, resulting in small absorption. In the bottom panel, the momentum distribution is convoluted with the experimental resolution and compared to the experimental data in \cite{Kubota2020}, finding a very good agreement. In \cite{Kubota2020}, a very similar procedure was used but the weight of the different components of $^{11}$Li were fitted to describe the data. It should be noted that, in the present work, these weights are obtained from the three-body calculation and have not been fitted, apart from an overall scaling factor to match the magnitude of the experimental data.

\begin{figure}[t]
 \centering
 \includegraphics[width=0.9\linewidth]{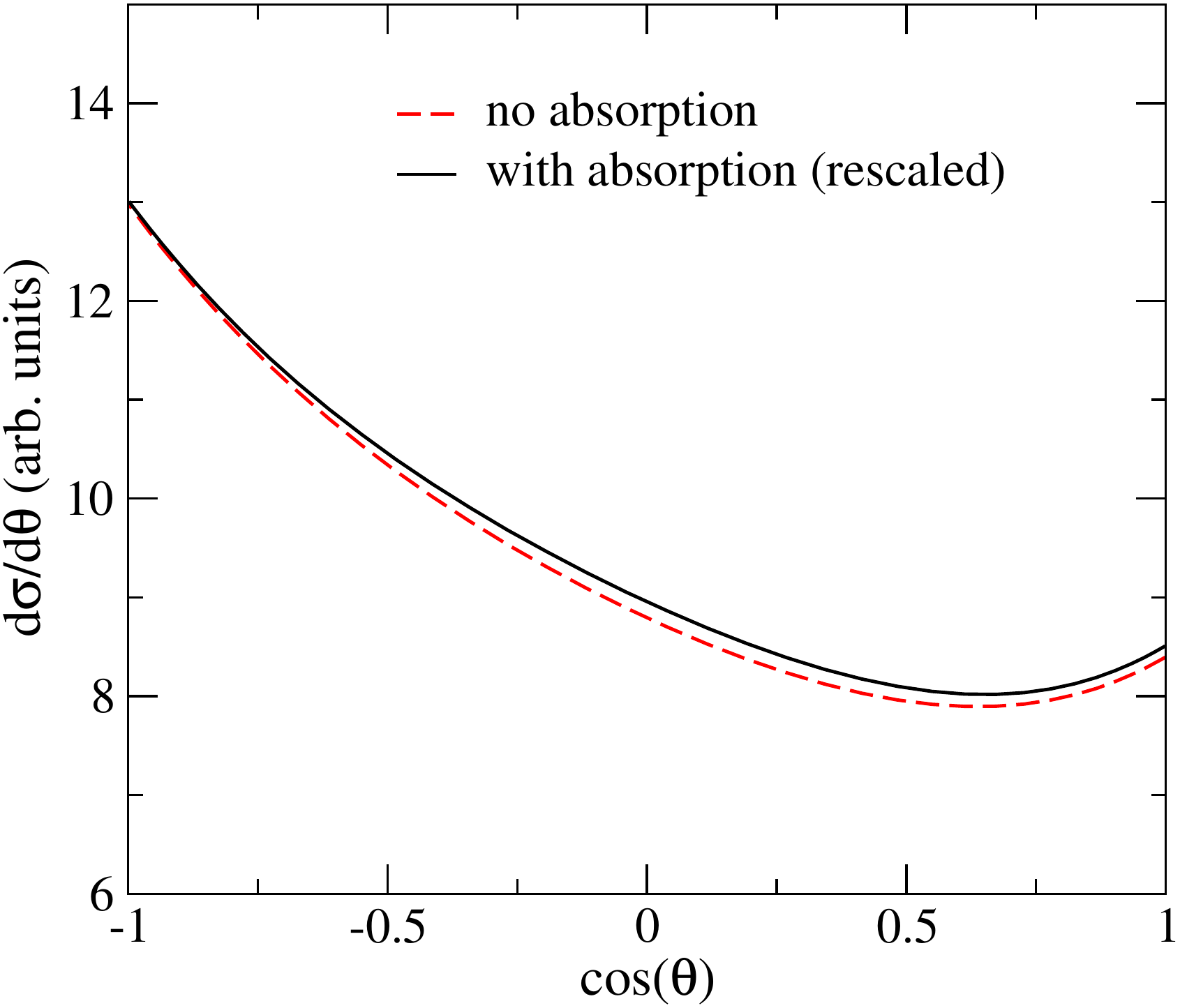}
 \caption{Opening angle distribution for the $^{11}$Li$(p,pn)$ reaction. Results with (solid line) and without (dashed line) the effect of absorption are presented.}
 \label{fig:oa_dist}
\end{figure}

We also present in Fig.~\ref{fig:oa_dist} the computed distribution in the opening angle for the calculation with and without the absorption effect. As noted in previous works \cite{Sim99}, the marked asymmetry of the distribution is a clear indicator of a strong interference between components of different parities, in this case the $s$ and $p$ waves. %For this observable the phase of $S(y)$, or, equivalently, the real part of the $V_{pC}$ potential, does play a role, in contrast to the missing momentum distribution, as the different components $w_\eta(k_x,k_y)$ in Eq.~(\ref{eq:49simp}) interfere. 
As can be seen in Fig.~\ref{fig:oa_dist}, the absorption effect keeps being rather moderate. This can be understood from the missing momentum distribution in Fig.~\ref{fig:kmiss} by noticing that the opening angle distribution results from the integration of Eq.~(\ref{eq:49simp}) over $k_x$ and $k_y$. From Fig.~\ref{fig:kmiss}, we see that the main contribution corresponds to a range of $k_y$ where the effects of the absorption are small, so this translates to a reduced effect in the opening angle. On the other hand, the $d$-wave, despite having a small contribution in the missing momentum distribution, plays a fundamental role in the shape of the opening angle distribution. From Eq.~(\ref{eq:49simp}), the opening angle distribution is a sum of Legendre polynomials restricted by the orbital angular momentum transfer between the different components of the $^{11}$Li ground-state wavefunction. Had the wavefunction only $s_{1/2}$ and $p_{1/2}$ components, $L$ would be restricted to 0 and 1, resulting in a linear shape for the opening angle distribution. It is the addition of the $d$-wave component which introduces higher multipoles and a curved shape to the distribution, despite its small magnitude. Similar effects were found for $^{11}$Li breakup on a carbon target explored with a 3-body description of $^{11}$Li \cite{Gar02}. The effect of the $f$-wave, not considered here, has been tested and found to be negligible.

\begin{figure}[t]
 \centering
 \includegraphics[width=0.9\linewidth]{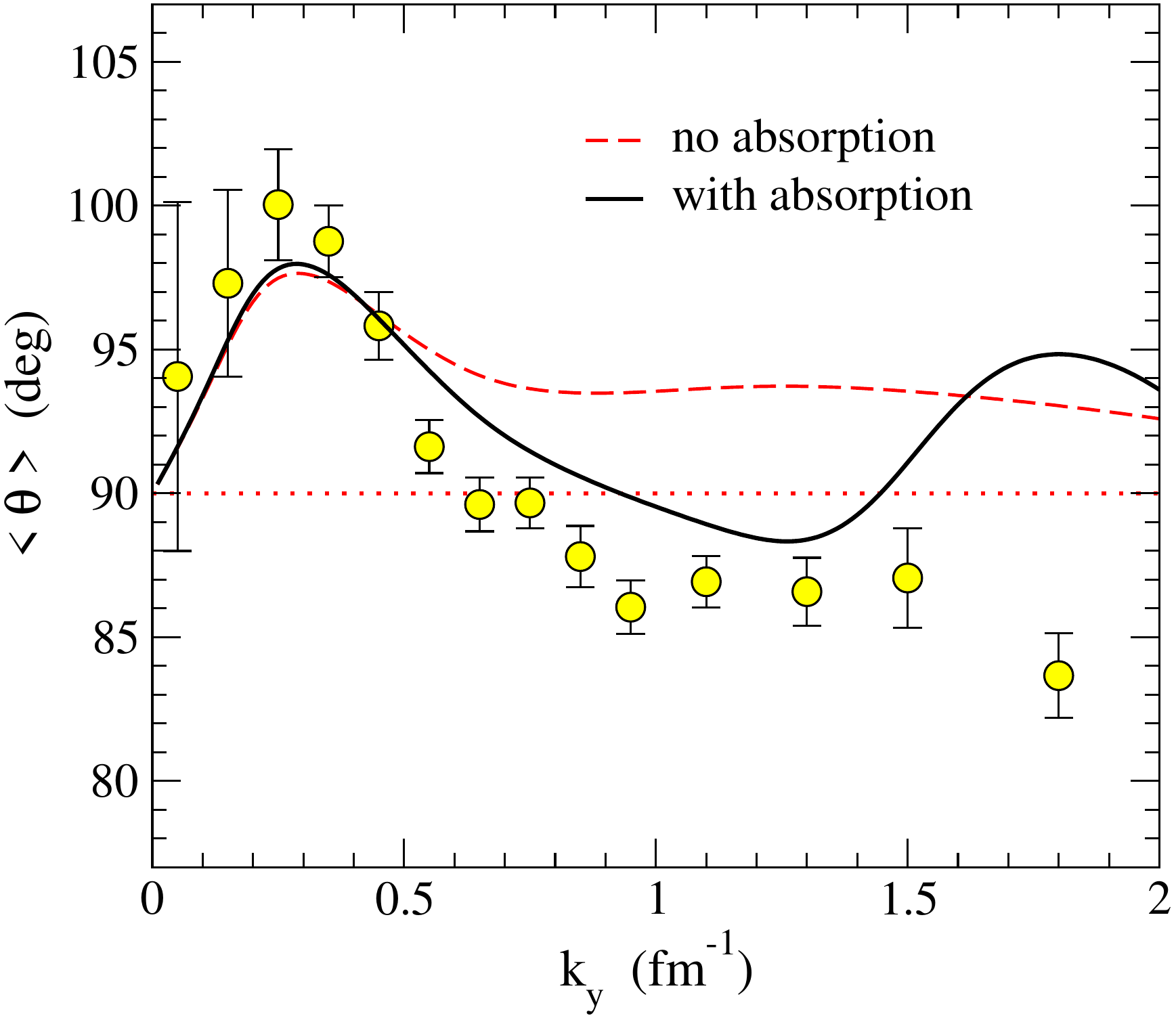}
 \caption{Average opening angle $\theta$ as a function of $k_y$. Results with and without absorption are presented in solid and dashed lines respectively.}
 \label{fig:corr1}
\end{figure}

In Fig.~\ref{fig:corr1}, we present the average opening angle $\theta$ as a function of the missing momentum $k_y$ and compare to the experimental data from \cite{Kubota2020}. Results with and without absorption are presented in solid and dashed lines, respectively. As can be  seen in the figure, for small $k_y$ the effect of absorption is small, and the agreement with the data is reasonable in that region. For larger $k_y$, the difference between both curves grows rapidly. The calculations including absorption decrease following the data, although they tend to overestimate the opening angle, and beyond $\approx 1.5$ fm$^{-1}$ the result presents an oscillation not seen in the data. Meanwhile, the calculations without absorption show a flat behaviour at $\approx 94^\circ$, far away from the experimental data.

\begin{figure}[t]
 \centering
 \includegraphics[width=0.9\linewidth]{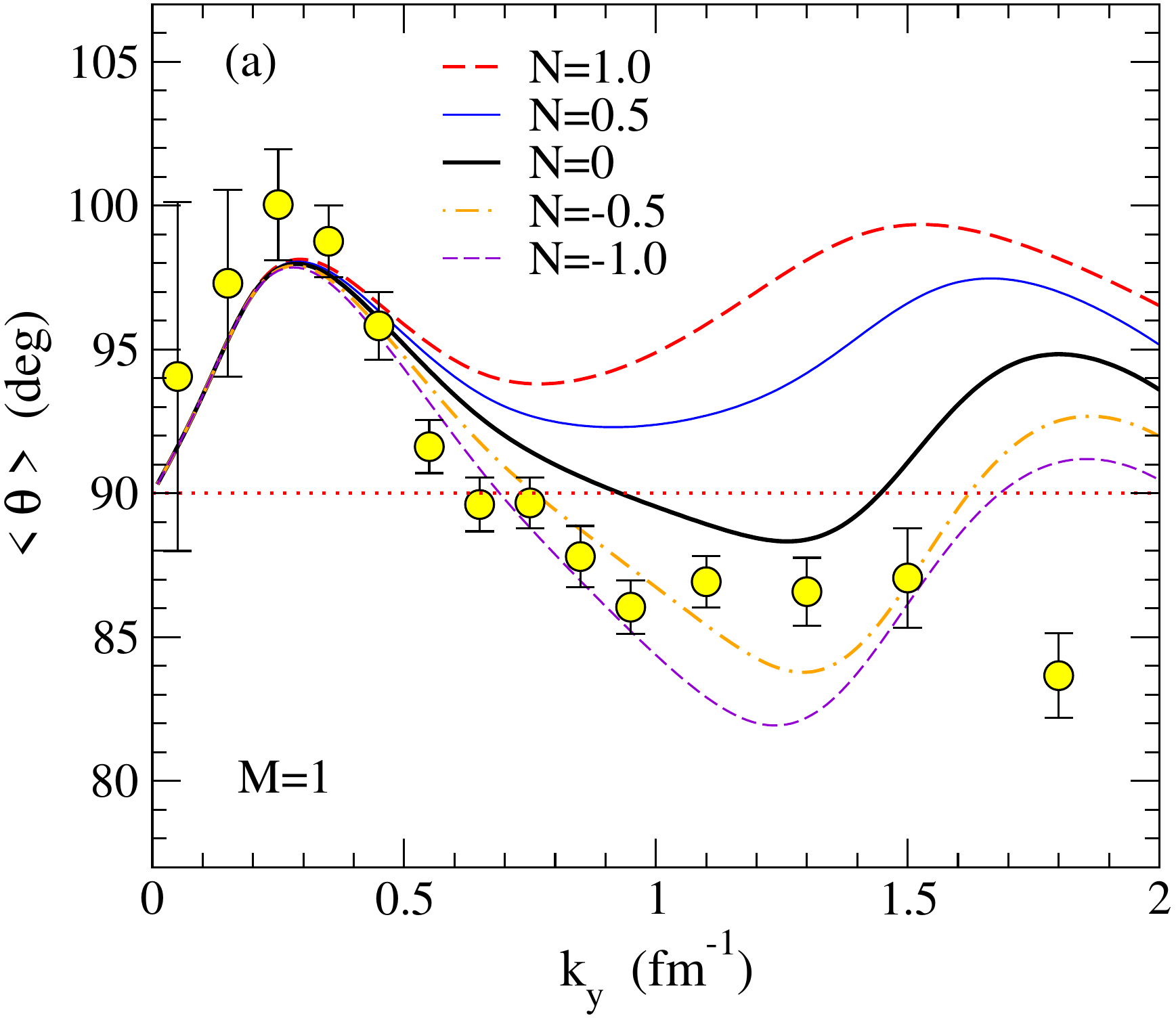}
 
  \includegraphics[width=0.9\linewidth]{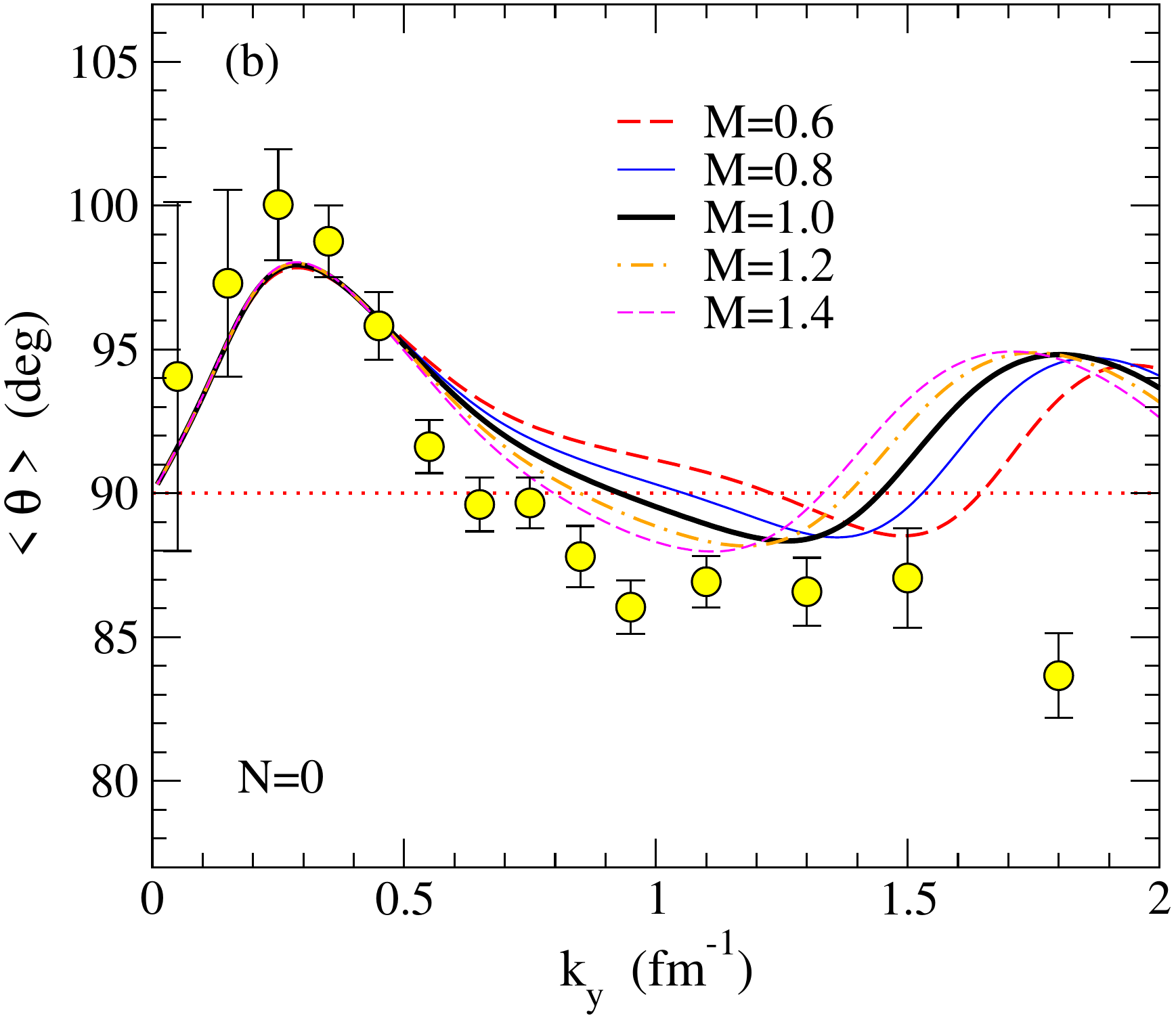}
 \caption{Effect of $S(y)$ on the average opening angle. $N$ denotes a scaling factor for the real part of the $V_{p\text{-}{^9\text{Li}}}$ potential, while $M$ corresponds to the scaling factor of the imaginary part. a) corresponds to $M=1$, that is, the original imaginary part of the potential, while b) corresponds to $N=0$, the removal of the real part of the potential.}
 \label{fig:corr3}
\end{figure}

Due to the simplicity of the reaction model, it is questionable whether the distortion effect due to the real part of the proton-target potential can be adequately described, in particular since the coordinates of the remaining neutron in $^{10}$Li are not considered (due to the lack of imaginary part in the $V_{pn}$ potential at these energies, this issue does not appear when considering absorption). We find it illustrative, however, to explore the sensitivity of the opening angle distribution to the real and imaginary parts of the potential (equivalently the phase and modulus of the $S$-matrix) to elucidate which parts of the distribution would be more affected by the reaction mechanism. We therefore modify the $p\text{-}^9$Li potential as follows 
\begin{equation}
V_{p\text{-}{^9\text{Li}}}(r)=NV(r)+iMW(r),
\end{equation}
i.e., we rescale the real part of the potential by a factor $N$ and we rescale the imaginary part by a factor $M$. For $S(y)$ this results in $S^\text{mod}(y)=|S(y)|^M e^{\textit{i}N\phi_S(y)}$, where the original $S(y)$ equals $S(y)=|S(y)|e^{\textit{i}\phi_S(y)}$. Previous calculations would correspond to $N=0$, $M=1$. The results of varying $N$ and $M$ are presented in Figure~\ref{fig:corr3}, where in the top panel the imaginary part of the potential is kept fixed to $M=1$ and the real part is modified, while in the bottom panel the real part of the potential is kept to zero and the imaginary part is modified. As can be seen in the figure, a variation of the real part of the proton-target potential leads to large modifications of the large-momentum behaviour of the opening angle distribution, while the low-momentum maximum is left mostly unchanged. With the original real part of the optical potential $N=1$, calculations deviate significantly from the experimental data, showing the limitations of the reaction model for the description of this distortion. For variations in the imaginary part of the optical potential, the maximum at low momenta remains unchanged as well, with larger modifications at large momenta, although the effect is more moderate than for the real part of the optical model. The fact that the low-momentum maximum is not modified by the proton-core potential can be associated to the halo nature of $^{11}$Li, where low $k_y$ momenta correspond to configurations where the two neutrons are far away from $^9$Li so the interaction $V_{p\text{-}{^9\text{Li}}}$ plays a small role. %In Fig.~\ref{fig:corr3}, we present the distribution for different values of $M$, the rescaling factor for the imaginary part, keeping the same real part ($N=1$, top panel) or removing it completely ($N=0$, bottom panel). As can be seen, the variation of both real and imaginary parts leads to significant differences in the distribution for large values of $k_y$, with the real part of the potential reducing the overall opening angle and the imaginary part introducing a somewhat oscillatory pattern in the distribution. On the other hand, the distribution at low values of $k_y$ is remarkably stable and, in particular, the position and magnitude of the maximum of the opening angle is left unchanged despite the rather large variations in the $V_{p^9\mathrm{Li}}$ potential. This can again be associated to the halo nature of $^{11}$Li, so low $k_y$ momenta correspond to configurations where the two neutrons are far away from $^9$Li so the interaction $V_{p^9\mathrm{Li}}$ plays a small role.

\begin{figure}[t]
 \centering
 \includegraphics[width=0.9\linewidth]{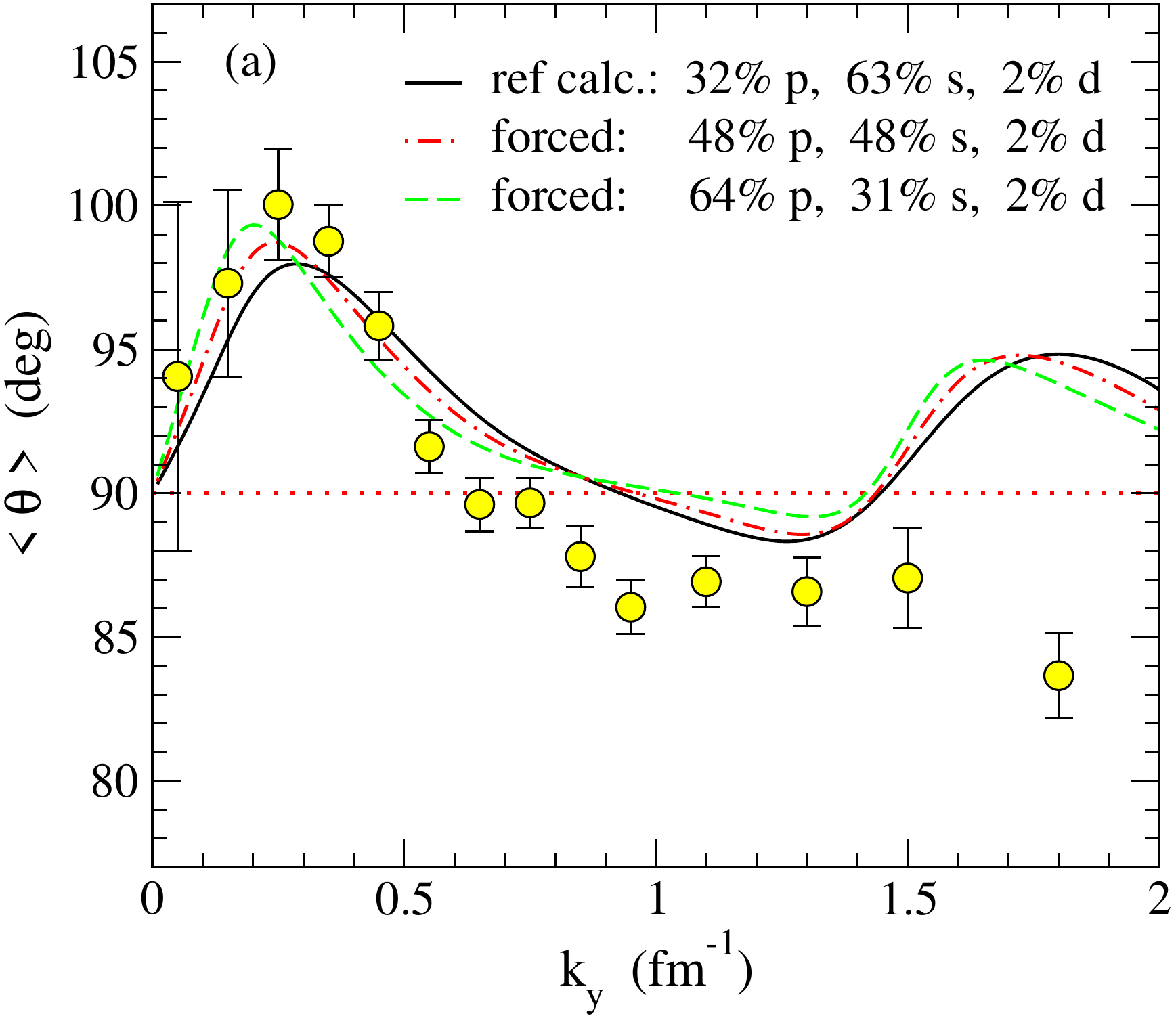}
 
  \includegraphics[width=0.9\linewidth]{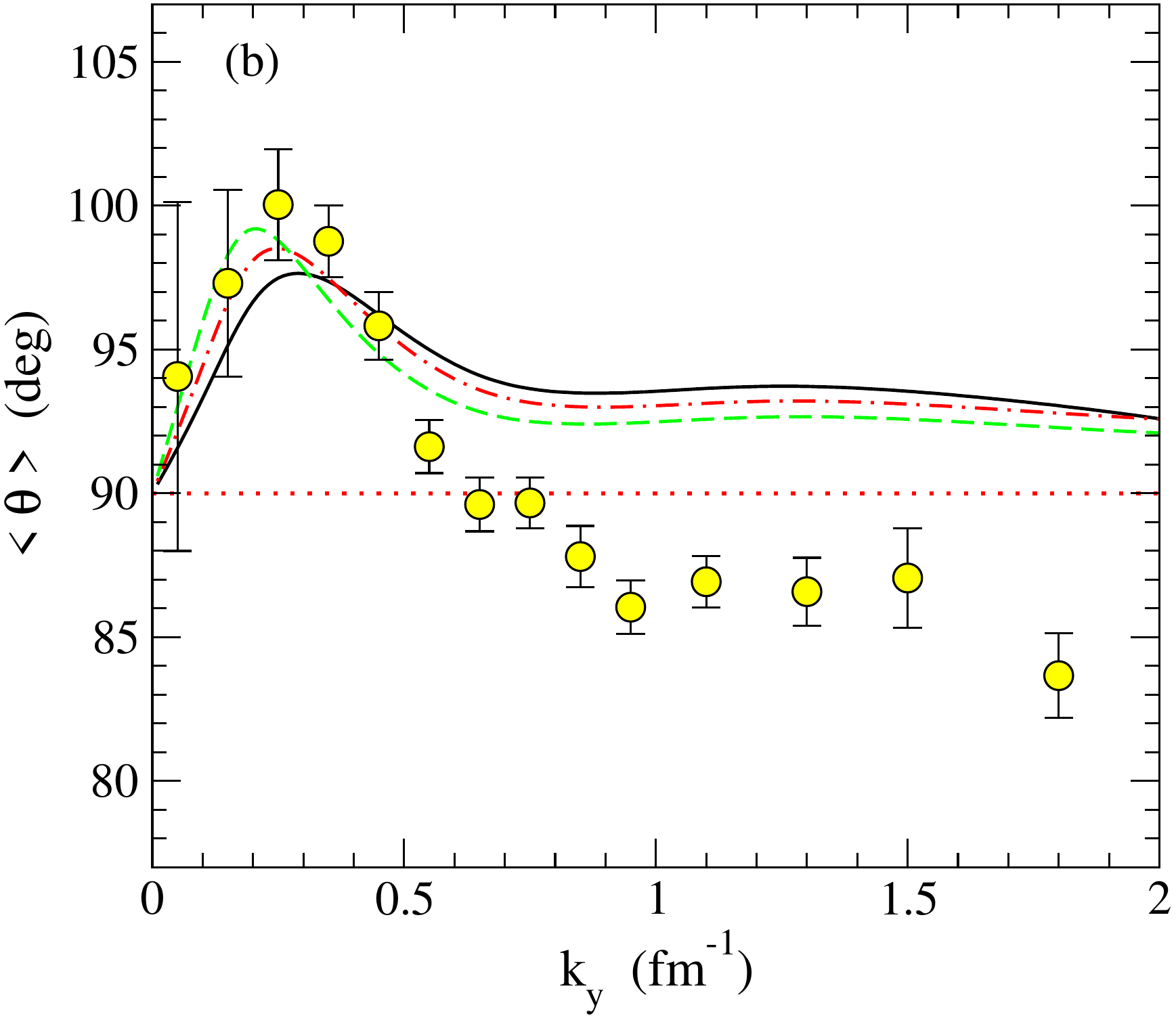}
 \caption{Average opening angle $\theta$ as a function of $k_y$, comparing the reference calculation (solid black line) with the results adjusting the $s$- and $p$-wave weights (dot-dashed red line and dashed green line), with (a) or without (b)  absorption.}
 \label{fig:corr2}
\end{figure}

Given that the large-momentum part shows such a large sensitivity to the description of the  $p\text{-}^9$Li interaction and thus to the reaction mechanism, we find it difficult to extract information on $^{10}$Li from this region of the distribution. As such, we choose to focus on the low-momentum maximum, finding that its magnitude is consistent but somehow underestimated by our calculations. Since the asymmetry of the opening angle distribution originates from the interference of the positive and negative parity waves, a larger asymmetry (and thus an opening angle more different from 90$^\circ$) should be obtained if the positive- and negative-parity waves have closer weights. Focusing on the region of the maximum, we can see from the momentum distribution in Fig.~\ref{fig:kmiss} that indeed the asymmetry originates from the interference of the $s$ and $p$ waves. As such, we have tried to increase the asymmetry by artificially increasing the contribution of the $p$ wave and reducing that of the $s$ wave so that they become more comparable. The results of this manipulation are presented in Fig.~\ref{fig:corr2}, the top (bottom) panel showing the results with (without) the effect of absorption. The original results, with 63\% $s$ wave and 32\% $p$ wave, are shown in the black solid line, while the weights for the red dash-dotted line and the green dashed lines are 48\% $s$, 48\% $p$ and 31\% $s$, 64\% $p$ respectively. As expected the results for the maximum are the same whether we consider absorption or not, although for larger momentum there is naturally a difference.

\begin{figure}[t]
 \centering
 \includegraphics[width=0.95\linewidth]{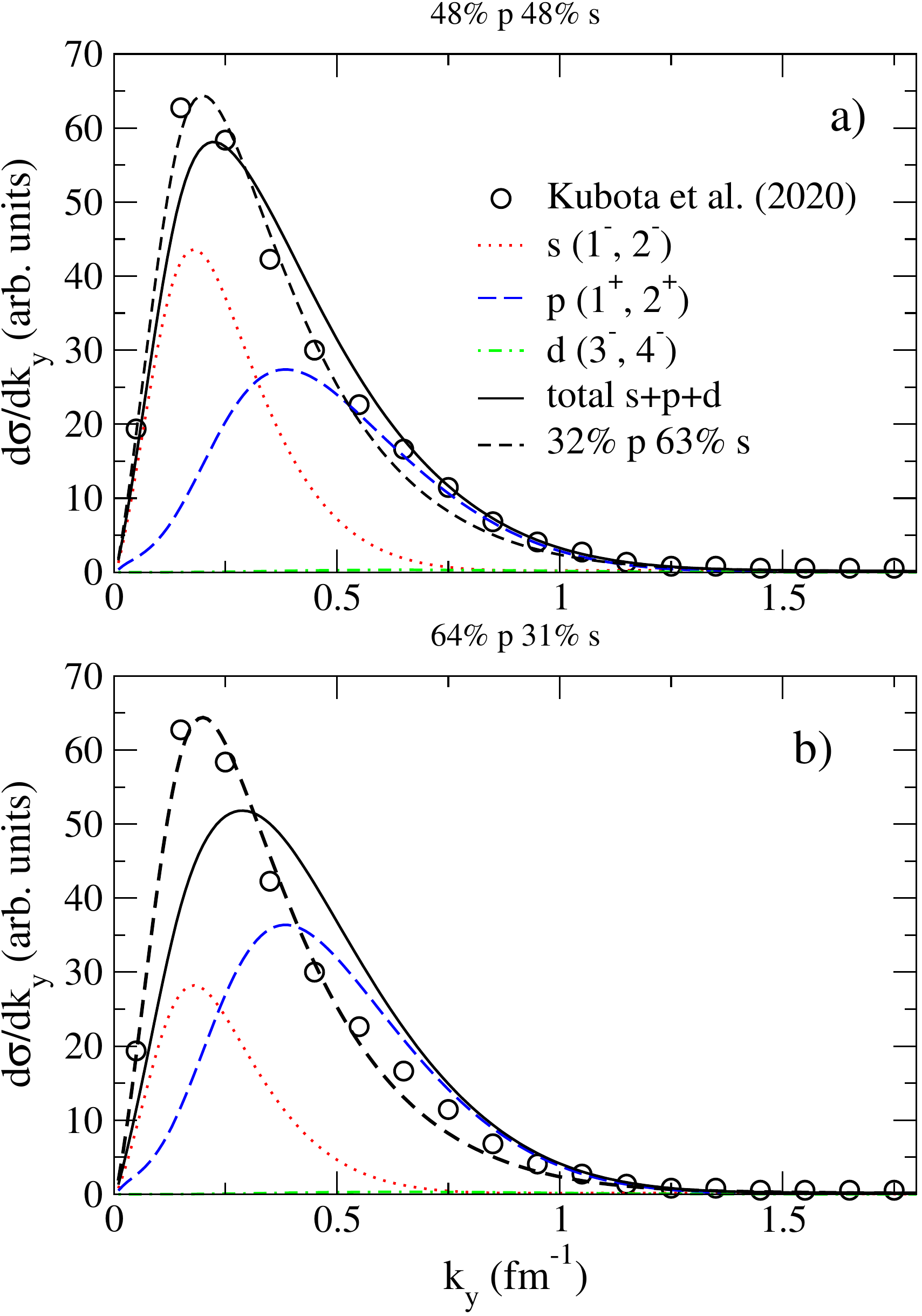}
 \caption{Momentum distribution described with the modified models with a) 48\% $s$ wave and 48\% $p$ wave and b)31\% $s$ wave and 64\% $p$ wave. The $s$, $p$ and $d$ components are shown separately. The distribution for the original calculation is shown in the black dashed line.}
 \label{fig:kmiss_mod}
\end{figure}

As expected, as well, a decrease in the $s$-wave component and an increase in $p$ wave does produce a larger opening angle, although for the more extreme modification the position of the maximum is also shifted and no longer fits that of the experimental data. Even despite this rather large alteration, the magnitude of the maximum in all cases is rather similar and smaller than the experimental data. By comparing the calculations to the experimental data in the maximum, it would seem the best agreement is found for the case with equal weight ($\approx$ 48\%) for the $s$ and $p$ waves, although it must be noted that the degree of agreement is comparable for all calculations. That $s$ and $p$ components have similar weights in the ground state of $^{11}$Li has also been suggested from other experiments \cite{Sim07,Sim99,Tan08}. In Fig.~\ref{fig:kmiss_mod} we show the momentum distributions with these modified weights. The different components are shown, as in Fig.~\ref{fig:kmiss}, alongside the original result (with weights 63\% and 32\% for the $s$ and $p$ wave, respectively) shown by the dashed black curve. The top panel corresponds to 48\% $s$ wave and 48\% $p$ wave and the bottom panel to 31\% $s$ wave and 64\% $p$ wave.

The sensitivity to the weight of each component is rather large in this observable, with the weights of 48\% $s$, 48\% $p$ describing well the large-momentum tail but missing the peak, and the 31\% $s$, 64\% $p$ missing the whole distribution. The degree of agreement for the top distribution is similar to that of the original distribution, so a clear distinction cannot be made from these observables for the weight of the two components, specially considering that the effect of absorption (as shown in Fig.~\ref{fig:kmiss}) modifies the agreement in a comparable manner. Given that the model with the original weights has been able to describe observables for $^{11}$Li$(p,pn)$ \cite{GomezRamos2017plb}, $^{11}$Li$(p,d)$ \cite{Casal2017plb} and $^{9}$Li$(d,p)$ \cite{Moro2019}, we favor the original weights. It is remarkable that the bottom distribution describes so poorly the data, given that the weights are similar to those extracted in \cite{Kubota2020}. The reason for this discrepancy could be related to the different descriptions of the structure of $^{11}$Li used here and in \cite{Kubota2020}. In particular, the scattering length found for the $2^-$ state is $-38$ fm in the model used in this work, while -45 fm \cite{kikuchi2013} for the one used in \cite{Kubota2020}. Therefore the distribution for the $s$ wave in this work is broader, so a description of the peak requires a larger $s$-wave component. It is then clear that the extraction of these structure properties from such observables is a model-dependent procedure, and a word of caution must be raised.

\vspace{-10pt}

\subsection{Sensitivity to the properties of the wavefunction: Comparison between $^{11}$Li and $^{19}$B}

\begin{figure}[t]
 \centering
 \includegraphics[width=0.9\linewidth]{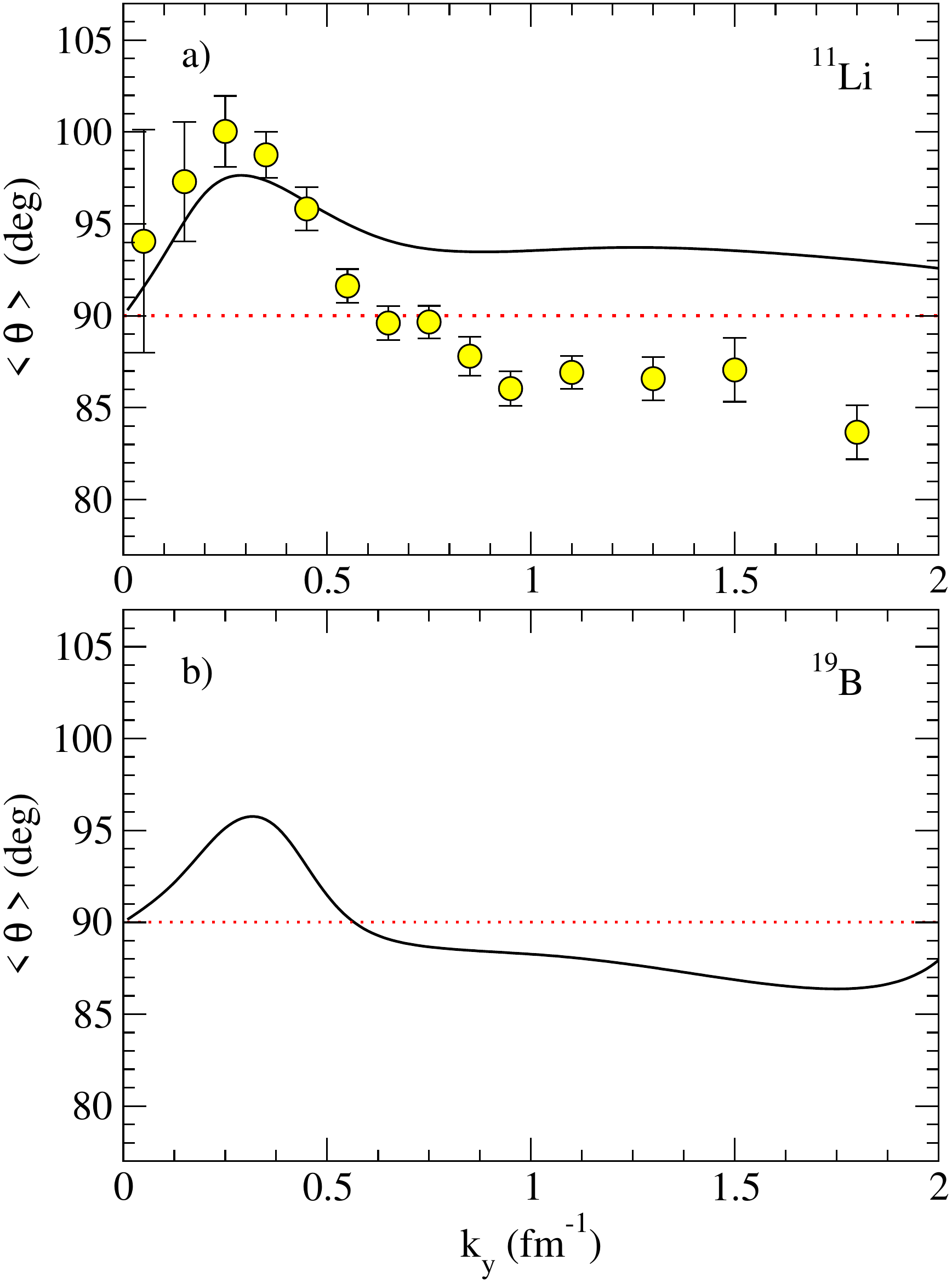}

 \caption{Average opening angle $\theta$ as a function of $k_y$ for a) $^{11}$Li and b) $^{19}$B. In both cases, absorption has been removed.}
 \label{fig:19b1}
\end{figure}

In this section, we will focus on the average opening angle as a function of the missing momentum and compare our $^{11}$Li results with the case of $^{19}$B, which was recently identified as a two-neutron halo via Coulomb dissociation~\cite{cook2020}. Given the dependence found for the large-momentum behaviour with absorption, we choose to focus on the low-momentum part of the distribution. We describe $^{19}$B using the $^{17}\text{B}+n+n$ model presented in Ref.~\cite{casal2020b19}, which for simplicity assumes an inert (and spinless) core. This model was shown to describe the observed $B(E1)$ distribution of $^{19}$B reasonably. The ground state is characterized by 53\% $s_{1/2}$ components and 39.2\% $d_{5/2}$ components, associated to a virtual state and a $d$-wave resonance in $^{18}$B, respectively, and with a very small admixture of $d_{3/2}$ and $p$ waves. As discussed in Ref. [15], better knowledge of the $^{18}$B spectrum would be required to include the core spin and effectively constrain the spin-spin interaction in $^{17}\text{B}+n$, similarly to our $^{11}$Li calculations in the previous section. In Fig.~\ref{fig:19b1}, we present the average opening angle as a function of missing momentum both for $^{11}$Li and $^{19}$B. 

\begin{figure}[t]
 \centering
 \includegraphics[width=0.9\linewidth]{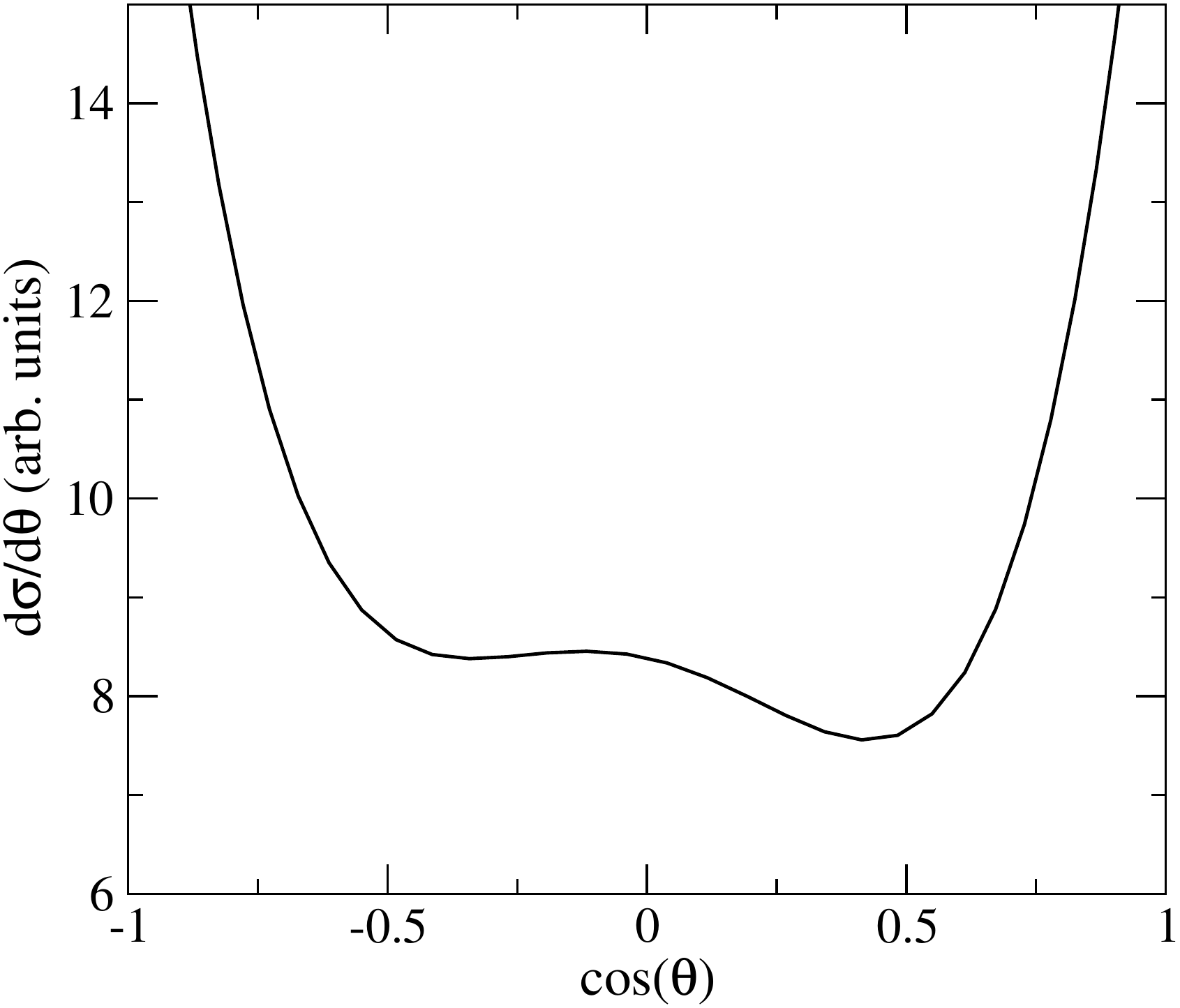}
 \caption{Opening angle distribution for the $^{19}$B$(p,pn)$ reaction. Absorption has not been included.}
 \label{fig:oa_distB}
\end{figure}

It is surprising that both distributions show very similar magnitudes at low momenta, since one would expect the magnitude of the opening angle to reflect the asymmetry of the wavefunction, and $^{19}$B, with only 3.3\% $p$-wave component, is markedly more symmetric (see Fig.~\ref{fig:oa_distB}) than $^{11}$Li. In fact, with the considered models, the overall opening angle for $^{11}$Li is of 95.4 degrees while for $^{19}$B it is of 91.7 degrees, much closer to the symmetric 90 degrees. This is consistent with the more asymmetric nature of $^{11}$Li, showing that the overall opening angle is a better descriptor of the asymmetry of the wavefunction than the maximum of the average angle as a function of the missing momentum. This can be understood noticing that the opening angle shown in Fig.~\ref{fig:19b1} has been averaged over the corresponding momentum range, so its magnitude will depend not only on the overlap between positive and negative parity components, as implied by Eq.~(\ref{eq:49simp}), but also on the magnitude of the wavefunction for that momentum range. Therefore, the interpretation of the magnitude of the maximum of the opening-angle distribution is not straightforward as it reflects not only the asymmetry of the wavefunction but also the overall magnitude of the wavefunction at the considered momentum.

\begin{figure}[t]
 \centering
 \includegraphics[width=0.95\linewidth]{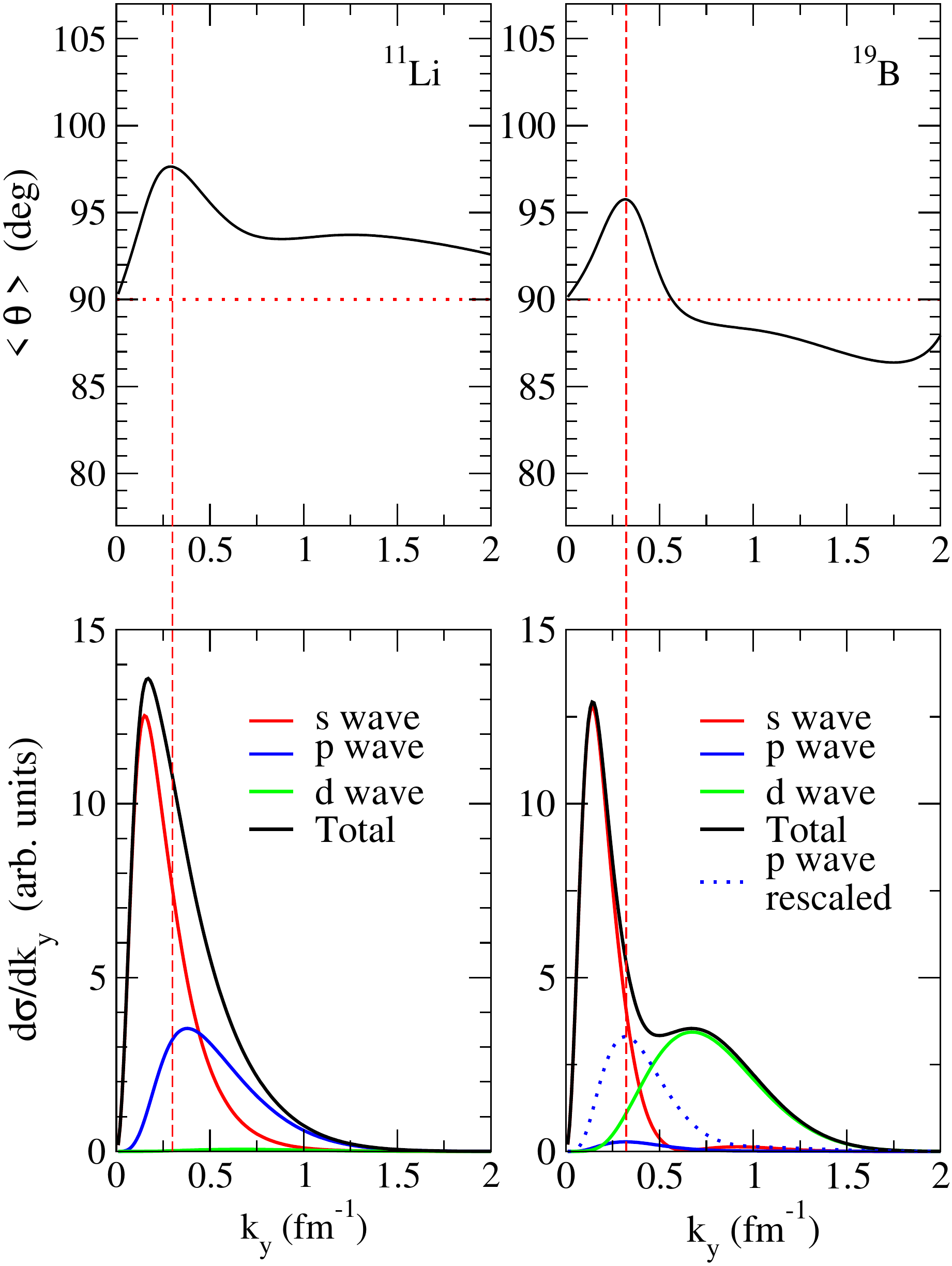}

 \caption{Average opening angle (top panels) and momentum distributions (bottom panels) for $^{11}$Li (left panels) and $^{19}$B (right panels). The different components are presented separately in the momentum distributions, as in Fig.~\ref{fig:kmiss}. A red dashed line is used for both nuclei to mark the position of the maximum in the average opening angle. In the momentum distribution for $^{19}$B (bottom right panel) the $p$-wave component is also presented rescaled for visibility by the dotted blue line.}
 \label{fig:19b2}
\end{figure}

On the other hand, the position of the maximum should indicate the value of the missing momentum for which the overlap between positive- and negative-parity components is maximal, when compared to the overall magnitude of the wavefunction for that momentum. To explore this, we present in Fig.~\ref{fig:19b2} the opening angle (top panels) as well as the momentum distributions (bottom panels) both for $^{11}$Li (left panels) and $^{19}$B (right panels). A red dashed line indicates the momentum with the maximum average opening angle. As can be seen in the figure, for $^{11}$Li, the maximum occurs for a momentum close to the maximum of the $p$ wave distribution, where $s$ and $p$ wave have similar magnitudes, as would be expected. The maximum average angle does not occur at the point where both components have exactly the same magnitude because their overlap not only depends on their magnitude but also their shape and phase, so a maximum overlap does not always occur for waves of the same magnitude.

For $^{19}$B, positive- and negative-parity components have very different magnitudes for all momenta, due to the $p$-wave component being Pauli blocked and contributing very little to the ground state. Here the same arguments for maximum overlap and similar magnitudes would predict that the maximal average opening angle would correspond to a momentum where either the larger component has a minimum or the smaller component has a maximum (with the caveats mentioned above). Indeed, as shown in Fig.~\ref{fig:19b2}, rescaled for visibility in the dotted blue line, the $p$-wave component exhibits a maximum exactly in the position of the maximum opening angle. Note that even though our $^{19}$B model does not include the spin of the core, the same qualitative conclusions would be expected when including the spin-spin splitting, provided the overall position and relative weight of the different single-particle configurations is maintained. A more quantitative analysis including the spin of the core explicitly requires further investigation.
It should also be noted that the $p$-wave component in this model presents no resonances and its shape corresponds to purely non-resonant continuum.

This suggests that the average opening angle as a function of the missing momentum could be used as a promising probe to explore components of the wavefunction with small contribution to the overall ground-state wavefunction of Borromean nuclei, if they have a parity opposite to that of the main components. This could help study the effect of core excitation or other structure properties that generate such components. For instance, the method could be used to study recent $(p,pn)$ data for $^{17}$B~\cite{Yang2021}, for which the analysis of the relative-energy spectrum and momentum distributions yielded mostly $s$ and $d$ waves. A deviation from 90 degrees in the corresponding average opening angle could then be interpreted as a result of a small $p$-wave admixture.

\subsection{Dineutron in coordinate space}\label{sec:dineutron}
\begin{figure}[t]
 \centering
 \includegraphics[width=0.9\linewidth]{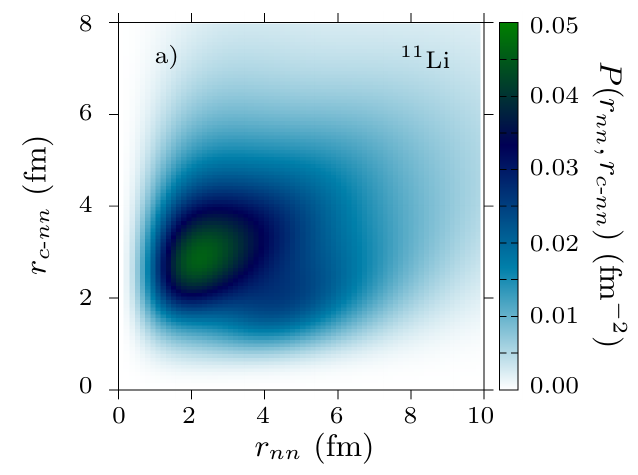}
 
 \includegraphics[width=0.9\linewidth]{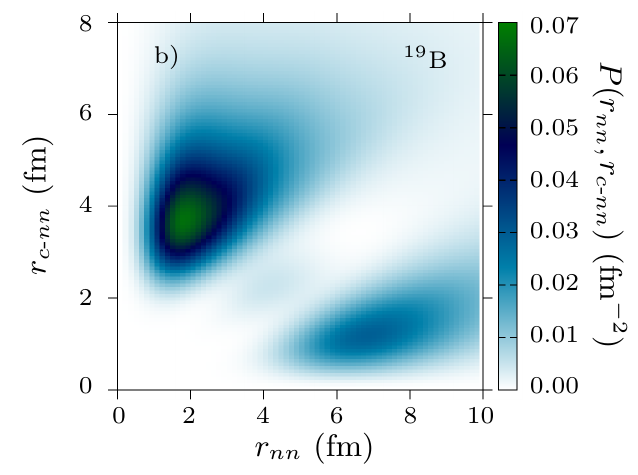}

 \caption{Ground-state probability density for a) $^{11}$Li and b) $^{19}$B as a function of $r_{nn}$ and $r_{c\text{-}nn}$.}
 \label{fig:xyprob}
\end{figure}

In Ref.~\cite{Kubota2020}, the maximum ($> 90$ deg.) of the average opening angle at low missing momenta is associated with the surface localization of dineutron correlations in $^{11}$Li. This means that, in coordinate space, the wavefunction favors configurations in which the valence neutrons are close to each other at some distance from the core. We can study this effect by looking at the ground-state probability density as a function of the distance between the halo neutrons ($r_{nn}$) and that between the center of mass of the two-neutron system and the core ($r_{c\text{-}nn}$). Here we use the usual Jacobi coordinates in the so-called $T$ representation. Our results for $^{11}$Li and $^{19}$B are shown in Fig.~\ref{fig:xyprob}.

The $^{11}$Li density (top panel) exhibits a clear maximum for small $n$-$n$ relative distances, the so-called dineutron configuration. Interestingly, the region corresponding to large $r_{nn}$ and small $r_{c\text{-}nn}$ values (sometimes called ``cigar''-like configuration) shows a small probability in our model. This is consistent with the large asymmetry of the opening angle in momentum space presented in Fig.~\ref{fig:opangle}, which is also apparent in coordinate space but now pointing towards small angles ($< 90$ deg.) between the valence neutrons. The r.m.s.~$r_{nn}$ as a function of $r_{c\text{-}nn}$ has a minimum at 3.13 fm, which is similar to the value of 3.2 fm given in Ref.~\cite{Hagino07}. It is clear from the present calculations that the wavefunction shows a diffuse tail at longer distances, and the dineutron peak appears in the surface. The conclusions are aligned with those in Ref.~\cite{Kubota2020}, though the authors obtained a value of $r_{c\text{-}nn}=3.6$ fm. Note, however, that this quantity is model dependent, as it comes from the three-body calculations, and it is not a direct result from the experimental data on the average opening angle. 

In the case of $^{19}$B (bottom panel of Fig.~\ref{fig:xyprob}), the corresponding density explores in general larger distances, with the minimum of the r.m.s.~$r_{nn}$ distance appearing at $r_{c\text{-}nn}=4.16$ fm. While a dineutron peak is also present, the wavefunction has a large probability outside this maximum. In particular, the ``cigar''-like configuration is more clearly separated and takes a big portion of the total norm. This is reflected by an overall $\sqrt{\langle r_{nn}^2\rangle}_{^{19}\text{B}}=3.65$ fm, which is significantly larger that that for $^{11}$Li, $\sqrt{\langle r_{nn}^2\rangle}_{^{11}\text{Li}}=3.30$ fm. Therefore, in $^{19}$B the ``dineutron'' is less compact, and this is consistent with a relatively small asymmetry in the opening angle, as discussed in the previous section. It is worth stressing again that the position of the maximum of the average opening angle as a function of the missing momentum is not directly related to the localization of the dineutron correlation in $r_{c\text{-}nn}$, and that a model is required to extract this information from the data.

\section{Summary and outlook}
\label{sec:conclusions}

In this work, we have analyzed the opening angle as a function of the missing momentum in nucleon-knockout reactions from Borromean two-neutron halo nuclei with proton targets. We have studied its sensitivity to the reaction mechanism through the $S(y)$ proton-core $S$-matrix and to the structure of the nucleus, in particular the weights of its positive- and negative-parity components. We have found a rather large effect of $S(y)$ for large missing momenta but an almost negligible one for small momenta. This indicates that caution should be taken when extracting structure information from the behaviour of this observable at large missing momenta, where the effects of the reaction mechanism are more pronounced. We have analyzed the data recently published for $^{11}$Li~\cite{Kubota2020} and found a reasonable agreement using the three-body model presented in Refs.~\cite{Casal2017plb,GomezRamos2017plb}, despite its $s$ and $p$-wave contributions being rather different from those extracted in Ref.~\cite{Kubota2020}. This shows a significant model dependence in the extraction of these quantities. We have also explored the $^{19}$B case, where even a small $p$-wave component produces significant variations in the opening-angle. This opens this observable as a sensitive probe to small components of the wavefunction with inverse parity to that of the dominant components, thus allowing their characterization, which can be difficult to explore otherwise. 
The surface localization of the dineutron explored in Ref.~\cite{Kubota2020} has been found for both of the nuclei considered in this work, although its relation to the opening angle as a function of the missing momentum requires an intermediate three-body model and is as such model-dependent. 

Improvements of the reaction theory and the structure description may allow for a better exploration of this observable. On the one hand, reaction calculations such as full distorted-wave impulse approximation (DWIA) \cite{Kaw18} or Transfer to the Continuum \cite{Gom18} would be desirable, but due to the numerous final states of the unbound $\text{core} + n$ system which have to be added coherently, this can pose a challenging and computationally expensive endeavour. Therefore, an improvement of the present theory, such as a description of $S(\boldsymbol{b}_y)$ in its natural coordinate (the impact parameter), the inclusion of distortion effects distinguishing explicitly the neutron and core, together with a more microscopic description of the proton-core interaction, may provide a better description of the large momentum part, which can give new insights on the properties of Borromean two-neutron halos. On the other hand, a refinement of three-body structure models, such as the inclusion of core excitations explicitly, may modify the results and conclusions. Work along these lines is ongoing and will be the subject of future research.

%%%%%%CORE EXCITATION

\begin{acknowledgments}
The authors would like to thank A. M. Moro, A. Corsi and Y. Kikuchi for useful discussions. This work has been partially supported by the Spanish Ministerio de Ciencia, Innovación y Universidades under project No.~FIS2017-88410-P, by the European Union research and innovation programme under Marie Sk\l{}odowska Curie Actions, grant agreement 101023609, and by the European Social Fund and Junta de Andalucía (PAIDI 2020) under grant number DOC-01006. 
\end{acknowledgments}

\appendix
\section{Three-body hyperspherical formalism for $\text{core}+n+n$ nuclei}

\begin{figure}[t]
 \centering
 \includegraphics[width=0.6\linewidth]{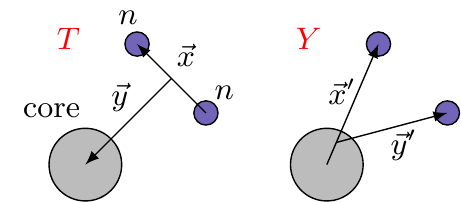}
 \caption{Jacobi-$T$ (left) and $Y$ (right) representations of a $\text{core}+n+n$ nucleus.}
 \label{fig:ttoy}
\end{figure}

We describe three-body wavefunctions using Jacobi coordinates and the hyperspherical framework, which is briefly discussed in this Appendix. A more comprehensive formulation is presented, for instance, in Refs.~\cite{zhukov93,nielsen01}, and summarized in Ref.~\cite{casal2020f29}. In the case of $\text{core}+n+n$ nuclei, two sets of Jacobi coordinates $\{\boldsymbol{x},\boldsymbol{y}\}$ can be defined, the $T$ and $Y$ representations, which are illustrated in Fig.~\ref{fig:ttoy}. The relation between the scaled Jacobi coordinates and the physical distances $\boldsymbol{r}_x,\boldsymbol{r}_y$ in the $T$ representation is given by
\begin{align}
    \boldsymbol{x}& =a_x \boldsymbol{r}_x, ~~a_x=\sqrt{1/2} \\
    \boldsymbol{y}& =a_y \boldsymbol{r}_y, ~~a_y=\sqrt{2A/(A+2)},
    \label{eq:jacobitor}
\end{align}
where $A$ is the mass of the core. From Jacobi coordinates, one defines the hyperspherical coordinates $\{\rho,\alpha,\widehat{x},\widehat{y}\}$, where $\rho=\sqrt{x^2+y^2}$ is the hyper-radius and $\alpha=\arctan{(y/x)}$ is the hyper-angle. 
Following the coupling order of Ref.~\cite{face}, and introducing $\Omega=\{\alpha,\widehat{x},\widehat{y}\}$, the wavefunctions for a given total angular momentum $j$ can be written as
\begin{align}
    \label{eq:HH1} 
  \psi^{j\mu}(\rho,\Omega) & = \rho^{-5/2}\sum_{\beta}U_{\beta}^{j}(\rho)\mathcal{Y}_{\beta}^{j\mu}(\Omega), \\
    \label{eq:HH2}
  \mathcal{Y}_{\beta}^{j\mu}(\Omega)&
  =\left\{\left[\Upsilon_{Kl}^{l_xl_y}(\Omega)\otimes\kappa_{S_x}\right]_{j_{ab}}\otimes\chi_I\right\}_{j\mu}, \\
    \label{eq:HH3}
  \Upsilon_{Klm_l}^{l_xl_y}(\Omega)& =\varphi_K^{l_xl_y}(\alpha)\left[Y_{l_x}(\boldsymbol{x})\otimes Y_{l_y}(\boldsymbol{y})\right]_{lm_l}, 
\end{align}
where $\beta\equiv\{K,l_x,l_x,l,S_x,j_{ab}\}$ are the relevant quantum numbers that label different components. Here, $\Upsilon_{Klm_l}^{l_xl_y}$ are the hyperspherical harmonics, eigenfunctions of the hypermomentum operator (or grand hyperangular momentum) $\widehat{K}$, and $\varphi_K^{l_xl_y}$ are the analytical hyperangular functions. Note that $S_x$ is the coupled spin of the two particles related by $x$, and $I$ is the spin of the core. Therefore, it is convenient to solve the problem and obtain the wavefunctions in the Jacobi-$T$ representation, with the two neutrons related by $x$. In that case, $S_x=0,1$ and the Pauli principle for two identical fermions imposes that $S_x+l_x$ is even. This reduces the number of components needed in the wavefunction expansion. The hyperradial functions $U_{\beta}^{j}$ are the solutions of a set of coupled equations with an effective three-body barrier defined by $K$ and involving the coupling potentials
\begin{equation}
V_{\beta'\beta}^{j\mu}(\rho)=\left\langle \mathcal{Y}_{\beta }^{j\mu}(\Omega)\Big|V_{12}+V_{13}+V_{23} \Big|\mathcal{Y}_{\beta'}^{ j\mu}(\Omega) \right\rangle +\delta_{\beta\beta'}V_{3b}(\rho).
\label{eq:3bcoup}
\end{equation}
In this expression, $V_{ij}$ are the pairwise potentials, usually adjusted to reproduce known properties of the binary subsystems, and $V_{3b}$ is a diagonal three-body force that is customarily used to fine-tune the computed three-body energies to the experimental values. In this work, instead of solving the coupled hyperradial equations numerically, the eigenstates of the three-body system are obtained by diagonalizing the three-body Hamiltonian in a discrete basis. We choose here, as in Refs.~\cite{casal2020b19,casal2020f29}, the analytical Transformed Harmonic Oscillator basis~\cite{casal13}. The convergence of the calculations is checked with the number of basis functions, but also setting a maximum hypermomentum $K_{max}$ for the wavefunction expansion. This limits the possible $l_x,l_y$ values, so no additional truncation is needed.

%The above expressions are valid in any Jacobi representation, provided one uses the appropriate spins.  
Once the solutions in the Jacobi-$T$ set are known, a transformation to the Jacobi-$Y$ representation can be achieved by using the Raynal-Revai coefficients~\cite{RR70}. Formally, the transformed wavefunction reads
\begin{align}
    \nonumber \psi_{Y}^{j\mu}(\rho,\Omega') & = \rho^{-5/2}\sum_{\beta'} \left(\sum_{\beta}N_{\beta,\beta'}U_{\beta}^{j}(\rho)\right) \mathcal{Y}_{\beta'}^{j\mu}(\Omega')\\
    & = \rho^{-5/2}\sum_{\beta'} \mathcal{U}_{\beta'}^j(\rho) \mathcal{Y}_{\beta'}^{j\mu}(\Omega'),
    \label{eq:wfY}
\end{align}
in terms of the quantum numbers in the $Y$ representation $\beta'$. The coefficients $N_{\beta,\beta'}=\langle Y,\beta'|T,\beta \rangle$ are analytical and determine the hyperradial functions $\mathcal{U}_{\beta'}^j$ in the new set. The angular functions in Eq.~(\ref{eq:wfY}) follow the same form of Eq.~(\ref{eq:HH2}) and with the same $K,l$, since these quantum numbers (as well as $\rho$) are preserved in the transformation~\cite{face}. Note that, the relation between Jacobi-$Y$ coordinates and physical distances is now given by
\begin{align}
    \boldsymbol{x}'& =a_{x'} \boldsymbol{r}_{x}', ~~a_{x'}=\sqrt{A/(A+1)} \\
    \boldsymbol{y}'& =a_{y'} \boldsymbol{r}_{y}', ~~a_{y'}=\sqrt{(A+1)/(A+2)},
    \label{eq:jacobitorY}
\end{align}

In Sec.~\ref{sec:theory}, we employ the three-body wavefunction in Jacobi-$Y$ coordinates and follow a different coupling order, which involves the single-particle angular momentum between one of the neutrons and the core. The corresponding radial functions in Eq.~(\ref{eq:wfcoord}) are given by Eq.~(A8) of Ref.~\cite{casal2020f29}, i.e.,
\begin{align}
    \nonumber  \omega_{\eta}^j(x',y')& =\hat{j}_1\hat{j}_2 \sum_{j_{ab}'} (-)^{j_{ab}-l_y-j_1} \hat{j}_{ab}' W(j_1l_y'js;j_{ab}'j_2) \\& \times \sum_{l} (-)^{l-l_x'-l_y'} \hat{l} W(l_y'l_x'j_{ab}'s;lj_1) \nonumber\\
    &\times \sum_{K}\rho^{-1/2} \mathcal{U}_{K\eta}^j(\rho)\varphi_{K}^{l_x'l_y'}(\alpha'),
    \label{eq:xyradch}
\end{align}
where $W$ are Racah coefficients, $\hat{\ell}=\sqrt{2\ell+1}$, and we have identified $\beta'=\{K,\eta\}$, with $\eta$ the set of quantum numbers introduced before.

\section{Wavefunction density in momentum space}

The derivation of the cross section in Sec.~\ref{sec:algebra} involves the density of a momentum-space wavefunction that incorporates final-state interaction and absorption effects, which is given in the Jacobi-$Y$ representation by Eq.~(\ref{eq:fourier1}). In order to obtain the angle between the two Jacobi vectors, we fix $\boldsymbol{k}_x$ in the direction of the $z$ axis by replacing the corresponding spherical harmonics by
\begin{equation}
    Y_{l_x,m_x} (0,0)=\frac{\hat{l_x}}{\sqrt{4 \pi}} \delta_{m_x,0}.
    \label{eq:Yx00}
\end{equation}
Then, by expanding the wavefunction couplings and performing the summation over all projections and unwanted angles, we get

\begin{align}
\nonumber&\dfrac{1}{\hat{j}^2}\sum_\mu\Phi_\text{gs}^*\Phi_\text{gs}  =  \sum_{\eta\eta'} w_\eta^*(k_x,k_y) w_{\eta'}(k_x,k_y) \hat{j_1}\hat{j_2}\hat{j_x}\hat{j_1'}\hat{j_2'}\hat{j_x'} \\
\nonumber & \times \dfrac{\hat{l_x}}{\sqrt{4 \pi}}\dfrac{\hat{l_x'}}{\sqrt{4 \pi}}  \frac{\hat{l_y}\hat{l}_y'}{\sqrt{4\pi}}\sum_{L}\hat{L}\tj{l_y}{l_y'}{L}{0}{0}{0}\tj{l_x}{l_x'}{L}{0}{0}{0}Y_{L0} (\theta) \nonumber  \\
\nonumber &\times\sj{j_x}{L}{j_x'}{j_1'}{I_c}{j_1}\sj{j_x}{L}{j_x'}{l_x'}{s_n}{l_x}\sj{j_2'}{j_2}{L}{j_1}{j_1'}{j}\sj{j_2'}{j_2}{L}{l_y}{l_y'}{I}.\\
\label{eq:etaend}  &\times(-)^{-2j_1'-I_c-s_n-I-j_x-j_x'+L-j_2'-j_2-j}.
\end{align}
By comparing this to Eq.~(\ref{eq:49simp}), we identify the geometrical quantities
\begin{align}
    \nonumber C_{\eta\eta'}& =\hat{j_1}\hat{j_2}\hat{j_x}\hat{j_1'}\hat{j_2'}\hat{j_x'} \dfrac{\hat{l_x}}{\sqrt{4 \pi}}\dfrac{\hat{l_x'}}{\sqrt{4 \pi}}  \frac{\hat{l_y}\hat{l}_y'}{\sqrt{4\pi}} \\
    & \times (-)^{-2j_1'-I_c-s_n-I-j_x-j_x'-j_2'-j_2-j},\label{eq:geom1}\\
    \nonumber D_{\eta\eta'}^{(L)} & =\hat{L}(-)^{L} \sj{j_x}{L}{j_x'}{j_1'}{I_c}{j_1}\sj{j_x}{L}{j_x'}{l_x'}{s_n}{l_x}\\
    & \times \sj{j_2'}{j_2}{L}{j_1}{j_1'}{j}\sj{j_2'}{j_2}{L}{l_y}{l_y'}{I}. \label{eq:geom2}
\end{align}

\bibliography{ref1}

\end{document}